\documentclass[11pt]{article}
\usepackage{graphicx}
\usepackage{amsmath}
\usepackage{amssymb}
\usepackage{amsxtra}

\usepackage{CJK}

\usepackage{graphicx,amsmath,amssymb}

\usepackage{graphicx}
 \usepackage{epstopdf}
\usepackage{amsmath}
\usepackage{dcolumn}
\usepackage{bm}
\usepackage{slashed}
\usepackage{siunitx}
\usepackage{ulem,xpatch}
\usepackage{hyperref}
\usepackage[mathlines]{lineno}
\usepackage{comment}
\usepackage{soul}
\usepackage{xcolor}
\usepackage{xfrac}
\hypersetup{colorlinks, linkcolor = [rgb]{0,0.0,0.5}, citecolor = [rgb]{0,0.0,0.5}, urlcolor = [rgb]{0,0.0,0.5}}

\def\Journal#1#2#3#4{{#1} {#2} (#4) #3 }

\def\PLB{{\em Phys. Lett.} B}

\def\PRL{\em Phys. Rev. Lett.}

\def\PRD{{\em Phys. Rev.} D}

\def\ZPC{{\em Z. Phys.} C}
\def\ZPA{{\em Z. Phys.} A}

\def\PPNP{\em Prog. Part. Nucl. Phys.}
\def\EPJC{{\em Eur. Phys. J.} C}

\def\ZP0{\em Zeit. Phys.}

\title{Advances for QCD and the Standard Model: Color-Confining Light-Front Holography and the Principle of Maximum Conformality}

\author{Stanley~J.~Brodsky\\SLAC National Accelerator Laboratory, Stanford University \\e-mail: sjbth@slac.stanford.edu}

\begin{document}
\maketitle

\begin{abstract}
I review how the application of superconformal quantum mechanics and light-front holography leads to new insights into the physics of color confinement,  the spectroscopy and dynamics of hadrons, as well as surprising supersymmetric relations between the masses of mesons, baryons, and tetraquarks.  Spontaneous chiral symmetry breaking is automatically fulfilled by supersymmetric Light-Front QCD. The light-front holographic approach (HLFQCD) also predicts the behavior of the QCD running coupling and other observables from the nonperturbative color-confining domain to the perturbative domain.   One can determine the QCD running coupling to high precision from the data of just a single experiment over the entire perturbative regime by using the Principle of Maximum Conformality (PMC).  The PMC, which generalizes the conventional Gell-Mann-Low method for scale-setting in perturbative QED to non-Abelian QCD, provides a rigorous method for achieving unambiguous scheme-independent, fixed-order Standard Model predictions,  consistent with the principles of the renormalization group. I also briefly review a novel feature of hadronic physics predicted by QCD: intrinsic heavy quarks.
\end{abstract}

\noindent QCD, Light-Front, Holography, Supersymmetry, Principle of Maximum Conformality, Intrinsic Quarks


\section{Color Confinement and Light-Front Holography}

A central problem in particle physics is to obtain a fundamental analytic description of hadrons, which not only has QCD color confinement, but also predicts the spectroscopy of hadrons and their light-front wave functions which underly their properties and dynamics. Guy de Teramond, Guenter Dosch, and I~\cite{Brodsky:2013ar}
have shown that a mass gap and a fundamental color confinement scale can be derived from light-front holography -- the duality between five-dimensional anti-de Sitter (AdS) space physical 3+1 spacetime using light-front time.  The combination of superconformal quantum mechanics~\cite{deAlfaro:1976je, Fubini:1984hf}, with light-front quantization~\cite{Dirac:1949cp} and the holographic embedding on a higher dimensional gravity theory~\cite{Maldacena:1997re} (gauge/gravity correspondence) has led to new analytic insights into the structure of hadrons and their dynamics~\cite{Brodsky:2013ar, deTeramond:2008ht, deTeramond:2014asa, Dosch:2015nwa, Brodsky:2014yha, Brodsky:2020ajy}. This novel approach to nonperturbative QCD dynamics, {\it holographic light-front QCD (HLFQCD}, has led to effective semi-classical relativistic bound-state equations for hadrons with arbitrary spin~\cite{deTeramond:2013it}, and it incorporates fundamental properties which are not apparent from the QCD Lagrangian, such as the emergence of a universal hadron mass scale, the prediction of a massless pion in the chiral limit, and remarkable connections between the spectroscopy of mesons, baryons and tetraquarks across the full hadron spectrum~\cite{Dosch:2015bca, Dosch:2016zdv, Nielsen:2018uyn, Nielsen:2018ytt}.

\section{Light-Front Theory and Holographic QCD}

When one observes an image in a flash photograph, each element is illuminated at a single light front time
 $\tau  = t + {z/c}$ along the front of the light wave. This is the basis of Dirac's "front form."
In contrast, each element of the image is illuminated at a different ``instant time" $t$.  Objects which are separated in space cannot interact at the same instant time, since information and interactions cannot travel faster than the speed of light.  This is the principle of causality.
Light-Front Hamiltonian theory~\cite{Dirac:1949cp} provides a causal, frame-independent, and ghost-free nonperturbative formalism for analyzing gauge theories such as QCD.

Measurements of hadron structure -- such as the structure functions determined by  deep inelastic lepton-proton scattering (DIS) -- are in fact analogous to a flash photograph: one observes the hadron at a fixed 
light-front time $ \tau =  t+z/c$ along a light-front.

The underlying physics is determined by
the hadronic light-front  wave functions (LFWFs)  
$\psi_n(x_i,  \vec k_{\perp i },  \lambda_i )$ with 
$x_i = {k^+_i\over P^+} = {k^0_i+k^z_i\over P^0+P^z}, \sum^n_i x_1 =1, \sum^n_i \vec k_{\perp _i} =\vec 0_\perp$  and spin projections $\lambda_i$.  The LFWFs are the Fock state projections of the eigenstates of the QCD LF invariant Hamiltonian $H_{LF} |\Psi\rangle = M^2|\Psi\rangle$~\cite{Brodsky:1997de}, where the LF Hamiltonian is the light-front time evolution operator defined directly from the QCD Lagrangian. 
One can avoid ghosts and longitudinal  gluonic degrees of freedom by choosing to work in the light-cone gauge  $A^+ =0$.  
The LFWFs are boost invariant; i.e.; they are independent of the hadron's momentum $P^+ =P^0 +P^z, \vec P_\perp.$
This contrasts with the wave functions defined at a fixed time $t$.  In fact,  the Lorentz boost of an instant-form wave function is much more complicated than a 
Melosh transform~\cite{Brodsky:1968ea} -- even the number of Fock constituents changes under a boost.
  
Current matrix elements such as form factors are simple overlaps of  the initial-state and final-state LFWFs, as given by the 
Drell-Yan West formula~\cite{Drell:1969km, West:1970av, Brodsky:1980zm}. 
Hadron form factors are matrix elements of the noninteracting electromagnetic current $j^\mu$  of  the hadron, as in the interaction picture of  quantum mechanics. One chooses the frame where the virtual photon 4-momentum $q^\mu$ has $q^+=0$, ${\vec q_\perp}^2=Q^2 = -q^2$ and $q^- P^+ = q\cdot p.$  One can also choose to evaluate matrix elements of $j^+= j^0+ j^z$ which eliminates matrix elements between Fock states with and extra $q \bar q$ pair.
There is no analogous formula for the instant form, since one must take into account the coupling of the external current to connected vacuum-induced currents.

Observables such as structure functions, transverse momentum distributions, and distribution amplitudes can all be defined from the hadronic LFWFs.  
The hadron distribution amplitudes $\phi_H(x_i, Q)$,  which enter factorization formulae for exclusive processes, are given by the valence LFWF integrated over transverse momentum $k^2_\perp < Q^2$. 

Since they are frame-independent, the structure functions measured in DIS are  the same whether they are measured at an electron-proton collider or in a fixed-target experiment where the proton is at rest.     There is no concept of length contraction of the hadron or nucleus at  a collider -- no collisions of  ``pancakes" --   since the observations  of the collisions of the composite hadrons are made at fixed light-front time $\tau$, not  at fixed time.    The dynamics of a hadron in the LF formalism are not dependent on the observer's Lorentz frame.

The frame-independent LF Heisenberg equation $H^{QCD}_{LF} |\psi_H\rangle = M^2_H \psi_H\rangle $ can be solved numerically by matrix diagonalization of the LF Hamiltonian in LF Fock space using ``Discretized Light-Cone  Quantization" (DLCQ)~\cite{Pauli:1985pv}, where anti-periodic boundary conditions in 
$x^-$ render the $k^+$ momenta  discrete  as well as  limiting the size of the Fock basis.  In fact, one can easily solve 1+1 quantum field theories such as QCD$(1+1)$ ~\cite{Hornbostel:1988fb} for any number of colors, flavors, and quark masses.  Unlike lattice gauge theory, the nonperturbative DLCQ analysis is in Minkowski space, it is frame-independent, and it is free of fermion-doubling problems.    A related method for solving nonperturbative QCD, ``Basis Light-Front Quantization" (BLFQ)~\cite{Vary:2009gt,Vary:2014tqa},  uses the eigensolutions of a color-confining approximation to QCD (such as LF holography ) as the basis functions, rather than the plane-wave basis used in DLCQ.  The LFWFs can also be determined from covariant Bethe-Salpeter wave function by integrating over $k^-$~\cite{Brodsky:2015aia}.  In fact, advanced quantum computers are now being used to obtain the DLCQ and BLFQ solutions.

Factorization theorems, as well as the DGLAP and ERBL evolution equations for structure functions and distribution amplitudes,  can be derived using the light-front Hamiltonian formalism~\cite{Lepage:1980fj}.  In the case of an electron-ion collider, one can represent the cross section for $e - p $ collisions as a convolution of the hadron and virtual photon structure functions times the subprocess cross-section in analogy to hadron-hadron collisions.   
This description of $\gamma^* p \to X$ reactions  provides new insights into electroproduction physics such as the dynamics of heavy quark-pair production, where 
intrinsic heavy quarks play an important role~\cite{Brodsky:2015uwa}.

In the case of $e p \to e^\prime X$, one can consider the collisions of the confining  QCD flux tube appearing between the $q$ and $\bar q$  of the virtual photon with the flux tube between the quark and diquark of the proton.   Since the $q \bar q$ plane is aligned with the scattered electron's plane, the resulting ``ridge"  of hadronic multiplicity produced from the $\gamma^* p$ collision will also be aligned with the scattering plane of the scattered electron.  The virtual photon's flux tube will also depend on the photon virtuality $Q^2$, as well as the flavor of the produced pair arising from $\gamma^* \to q \bar q$.  The resulting dynamics~\cite{Brodsky:2014hia} is a natural extension of the flux-tube collision description of the ridge produced in $p-p$ collisions~\cite{Bjorken:2013boa}.

\section{Other Features of Light-Front QCD}

There are a number of advantages if one uses  LF Hamiltonian methods for perturbative QCD calculations.  Unlike instant form, where one must sum  over $n !$ frame-dependent  amplitudes, only the $\tau$-ordered diagrams where every line has  positive $k^+ =k^0+k^z$  can contribute~\cite{Cruz-Santiago:2015dla}. The number of nonzero amplitudes is also greatly reduced by noting that the total angular momentum projection $J^z = \sum_i^{n-1 } L^z_i + \sum^n_i S^z_i$ and the total $P^+$ are  conserved at each vertex.  In addition, in a renormalizable theory the change in orbital angular momentum is limited to $\Delta L^z =0,\pm 1$ at each vertex.  The calculation of a subgraph of any order in pQCD only needs to be done once;  the result can be stored in a ``history" file, since in light-front perturbation theory,  the numerator algebra is independent of the process; the denominator changes, but only by a simple shift of the initial $P^-$. Loop integrations are three-dimensional: $\int d^2\vec k_\perp \int^1_0 dx.$
Renormalization can be done using the ``alternate denominator" method which defines the required subtraction counter-terms~\cite{Brodsky:1973kb}.

The LF vacuum in LF Hamiltonian theory is defined as the eigenstate of $H_{LF}$ with lowest invariant mass. Since propagation of particles with negative $k^+$  does not appear, there are no loop amplitudes appearing in the LF vacuum -- it is  is thus trivial up to possible $k^+=0$ ``zero"  modes.   The usual quark and gluon QCD vacuum condensates of the instant form =are replaced by physical effects,  such as the running quark mass and the physics contained within the hadronic LFWFs  in the hadronic domain. This is referred to as ``in-hadron" condensates~\cite{Casher:1974xd,Brodsky:2009zd,Brodsky:2010xf}.  In the case of the Higgs theory, the traditional Higgs vacuum expectation value (VEV) is replaced by a zero mode, analogous to a classical 
Stark or Zeeman field.~\cite{Srivastava:2002mw}   This approach contrasts with the traditional view of the vacuum  based on the instant form. 

The instant-form vacuum, the lowest energy eigenstate of the instant-form Hamiltonian,  is defined at one instant of time over all space; it is thus acausal and frame-dependent.  It is usually argued that the QCD contribution to the cosmological constant -- dark energy  -- is $10^{45}$ times larger that observed, and in the case of the Higgs theory, the Higgs VEV is argued to be $10^{54}$ larger than observed~\cite{Zee:2008zz}, 
estimates based on the loop diagrams of the acausal frame-dependent instant-form vacuum.  However, the physical universe is observed within the causal horizon, not at a single instant of time.  In contrast, the light-front vacuum provides a viable description of the visible universe~\cite{Brodsky:2010xf}. Thus, in agreement with Einstein's theory of general relativity, quantum effects do not contribute to the cosmological constant.   In the case of the Higgs theory, the Higgs zero mode has no energy density,  so again  it gives no contribution to the cosmological constant.  However, it is possible that if one solves the Higgs theory in a curved universe, the zero mode will be replaced with a field of nonzero curvature which could give a nonzero contribution.

\section{ Holographic light-front QCD}

A basic understanding of hadron
properties, such as confinement and the emergence of a mass scale, from
first principles in QCD has remained elusive: Hadronic characteristics are not explicit
properties of the QCD Lagrangian and perturbative QCD, so successful in the large
transverse momentum domain, is not applicable at large distances. This obstacle is overcome in holographic QCD with
the introduction of a superconformal symmetry in anti de Sitter (AdS) space which
is responsible for confinement and the introduction of a mass scale 
within the superconformal group. When mapped to light-front coordinates in physical spacetime,
this approach incorporates supersymmetric relations between the Regge trajectories
of meson, baryon and tetraquark states which can be visualized in terms of specific
SU(3) color representations of quarks.

Remarkably, LF theory in 3+1 physical space-time is holographically dual to five-dimensional AdS space, if one identifies the LF radial variable $\zeta$ with the fifth coordinate $z$ of AdS$_5$~\cite{Brodsky:2013ar, deTeramond:2008ht, deTeramond:2014asa, Dosch:2015nwa, Brodsky:2014yha, Brodsky:2020ajy}.   If the metric of the conformal AdS$_5$ theory is modified by  a dilaton of the form $e^{+ \kappa^2 z^2}$, one obtains an analytically-solvable  Lorentz-invariant color-confining LF Schr\"odinger equations for hadron physics.  The parameter $\kappa$ of the dilaton becomes the fundamental mass scale of QCD, underlying the color-confining potential of the LF Hamiltonian and the 
running coupling $\alpha_s(Q^2)$ in the nonperturbative domain.  When one introduces super-conformal algebra, the result is ``Holographic LF QCD" which not only predicts a unified Regge-spectroscopy of mesons, baryons, and tetraquarks, arranged as supersymmetric 4-plets, but also the hadronic LF wavefunctions which underly form factors, structure functions,  and  other dynamical phenomena.  In each case, the quarks and antiquarks cluster in hadrons as $3_C$ diquarks, so that mesons, baryons and tetraquarks all obey a two-body $3_C - \bar 3_C$ LF bound-state equation.  Thus tetraquarks are compact hadrons, as fundamental as mesons and baryons.
``Holographic LF QCD" also leads to novel phenomena such as the color transparency of hadrons produced in 
hard-exclusive reactions traversing a nuclear medium and asymmetric intrinsic heavy-quark distributions $Q(x) \ne \bar Q(x)$, appearing at high $x$ in the non-valence higher Fock states of hadrons~\cite{Brodsky:1980pb,Brodsky:1984nx}.

The light front holographic approach also incorporates
the essential consequence of spontaneous chiral symmetry breaking.   All of the typical features of spontaneous chiral symmetry breaking are automatically fulfilled by supersymmetric LFHQCD: There is a mass-
less boson (the pion in the chiral limit),  and the parity
doublets  $(\rho, A_1)$ and $(N, N(1535))$  have different masses.  
A detailed discussion is given in ref.~\cite{Dosch:2022rhj}

The holographic light-front QCD framework provides a unified nonperturbative description of the hadron mass spectrum, form factors and quark distributions.
In our article ~\cite{deTeramond:2021lxc}  we have extended our previous description of quark distributions \cite{deTeramond:2018ecg,Liu:2019vsn} 
in LF holographic QCD to predict 
the gluonic distributions of both the proton and pion from the coupling of the metric fluctuations induced by the spin-two Pomeron with the energy momentum tensor in anti-de Sitter space, together with constraints imposed by the Veneziano model without additional free parameters. The gluonic and quark distributions are shown to have significantly different effective QCD mass scales.
The comparison of our predictions with the gluon gravitational form factor computed from Euclidean lattice gauge theory and the gluon distribution in the proton and pion from global analyses also give very good results.

Phenomenological extensions of the holographic QCD approach  have also led to nontrivial connections between the dynamics of form factors and polarized and unpolarized quark distributions with pre-QCD nonperturbative approaches such as Regge theory and the Veneziano model~\cite{Sufian:2016hwn, deTeramond:2018ecg, Liu:2019vsn}. As discussed in the next section, it also predicts the analytic behavior of the QCD coupling $\alpha_s(Q^2)$ in the nonperturbative domain~\cite{Brodsky:2010ur, Deur:2014qfa}.

\section{Light-Front Holography QCD and Supersymmetric Features of Hadron Physics}  

\begin{figure}
 \begin{center}
\includegraphics[height= 10cm,width=15cm]{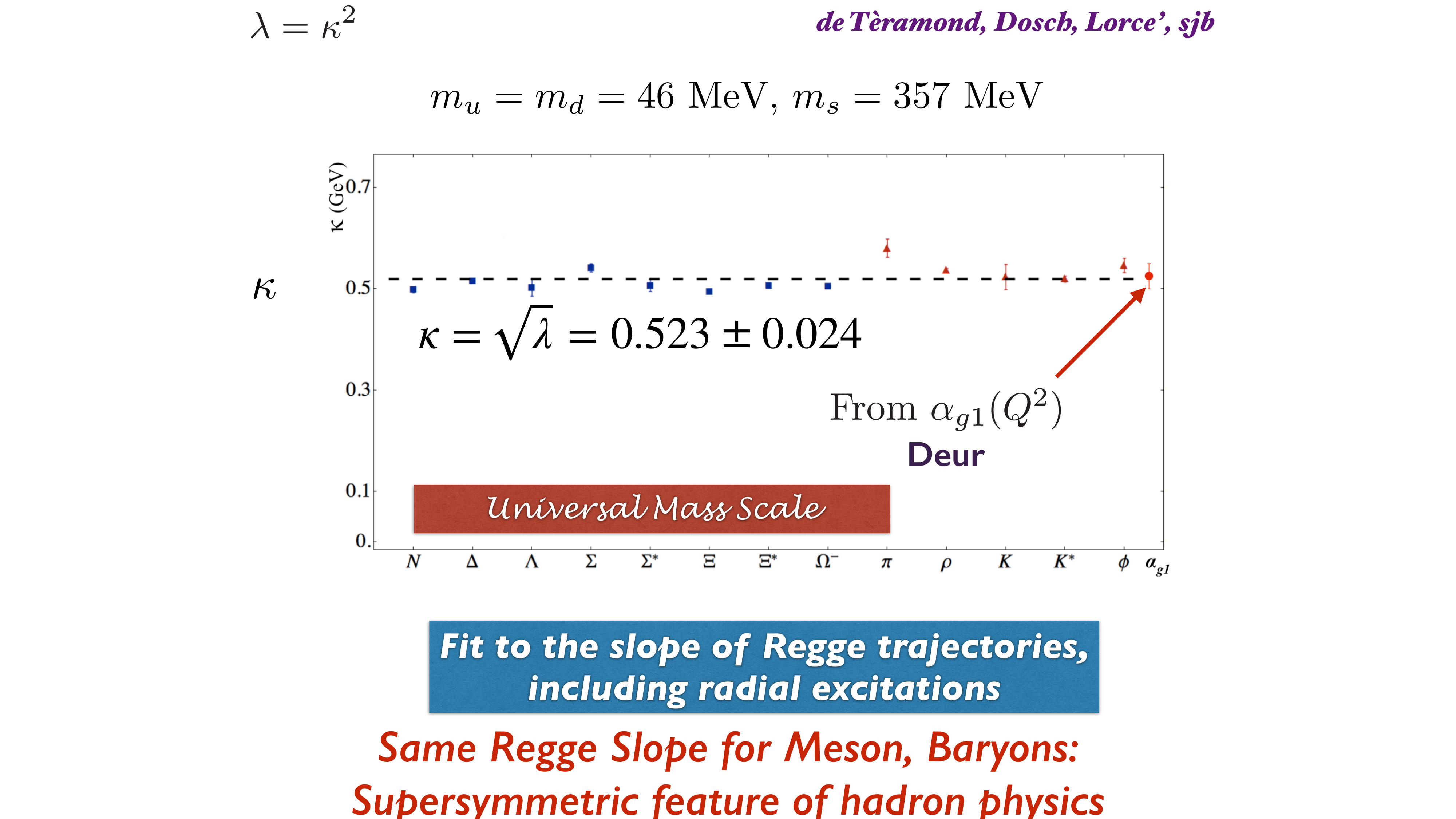}
\end{center}
\caption{The slopes of the measured meson and baryon Regge trajectories.}
\label{Bledslides2.pdf}
\end{figure}

\begin{figure}
 \begin{center}
\includegraphics[height= 9cm,width=18cm]{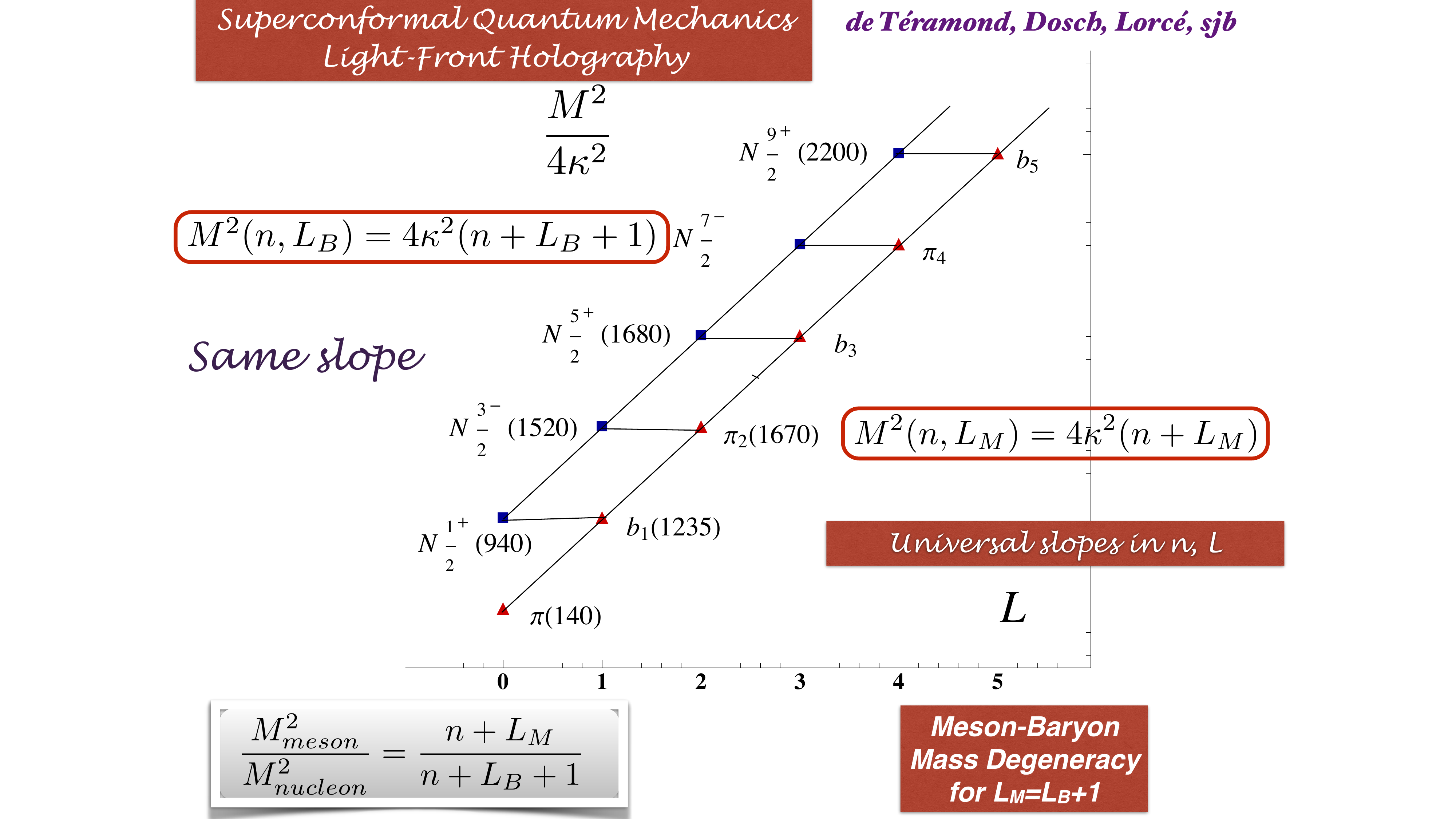}
\includegraphics[height=9 cm,width=15cm]{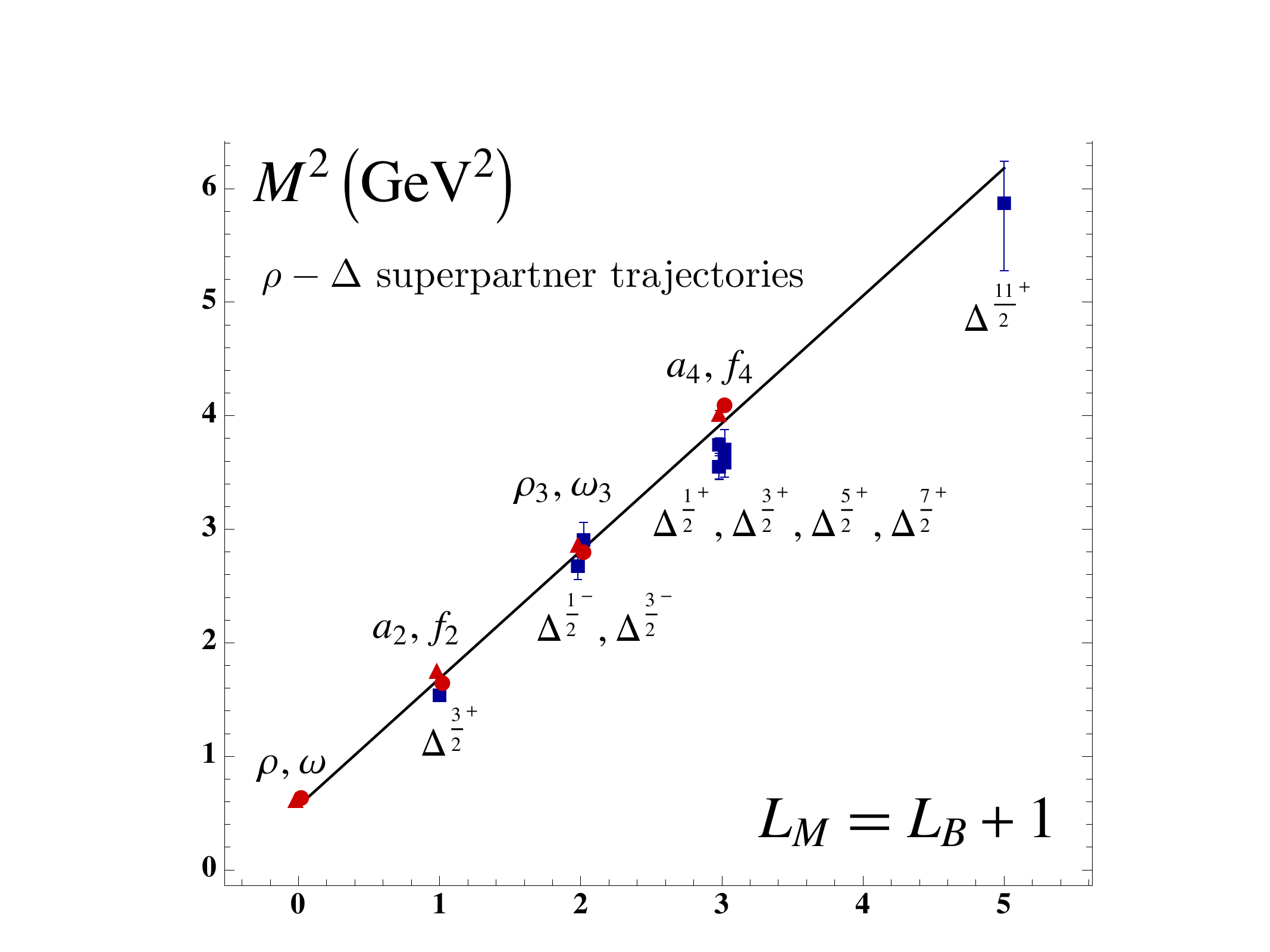}
\end{center}
\caption{Examples of supersymmetric meson and baryon Regge trajectories. Comparison of the pion  and proton trajectories and the comparison of the $\rho/\omega$ meson Regge trajectory with the $J=3/2$ $\Delta$  baryon trajectory.   The degeneracy of the  meson and baryon trajectories if one identifies a meson with internal orbital angular momentum $L_M$ with its superpartner baryon with $L_M = L_B+1$ using superconformal algebra.
See Refs.~\cite{deTeramond:2014asa,Dosch:2015nwa}.} 
\label{Bledslides8.pdf}
\end{figure}

One of the most remarkable feature of hadron spectroscopy is that, to a very good approximation, mesons and baryons are observed to lie on almost identical Regge trajectories:
$M^2_M = 2 \kappa^2 (n + L_M) $ 
for mesons with light quarks and  
$M^2_B = 2 \kappa^2 (n + L_B +1) $ 
for baryons with light quarks. 
The slopes $\lambda =\kappa^2$  in $M^2_H(n,L)$ are identical for both mesons and baryons in both the principal number $n$ and orbital angular momentum  $L$. 
(The index $n$ can be interpreted as the number of nodes in the resulting two-body wave function. )
The universality of the slopes of Regge trajectories across the hadronic spectrum is shown in Fig. \ref{Bledslides2.pdf}.  
An example comparing the pion and proton trajectories is shown in Fig. \ref{Bledslides8.pdf}.
This degeneracy between the Regge slopes of the two-body mesons and three-body baryons provides compelling evidence that two of the three quarks in the baryon valence Fock state pair up as 
diquark clusters. Then $L_M$ represents the orbital and angular momentum between the $3_C$ quark and  $\bar 3_C$ antiquark for mesons, and 
$L_B$ represents the orbital angular momentum between the $3_C$ quark and a $\bar 3_C$ spin-0 $[qq]$  or spin-1 $(qq)$ diquark in baryons.  
The identical $3_C- \bar 3_C$ color-confining interaction appears for mesons and baryon.  The index $n$ can be interpreted as the number of nodes in the resulting two-body wave function.

The unified spectroscopy of hadronic bosons and fermions point to an underlying {\it supersymmetry} between  
mesons and baryons in QCD.  In fact, the supersymmetric Light Front Holographic approach to QCD not only provides a unified spectroscopy of mesons and baryons, but it also predicts the existence and spectroscopy of tetraquarks:
the mass degeneracy of mesons and baryons with their tetraquark partners, bound states of $3_C$ diquarks and $\bar 3_C$ anti-diquarks.
The meson-baryon-tetraquark 4-plet predicted by the LF supersymmetric approach is illustrated in Fig.~5.  The baryon has two entries in the 4-plet,  analogous to the upper and lower spinor components of a Dirac spinor. 
For example, the  proton  $|[ud] u \rangle $ with $J^z=+1/2$ has equal probability to be a bound state of a scalar $[ud]$  diquark and a $u$ quark with $S^z=+1/2, L^z=0$ or the $u$ quark with nonzero orbital angular momentum $S^z=-1/2, L^z=+1$.  The spin-flip matrix element of the electromagnetic current  between these two states gives the proton's Pauli form factor in the light-front formalism~\cite{Brodsky:1980zm}.

\begin{figure}
 \begin{center}
\includegraphics[height= 8cm,width=15cm]{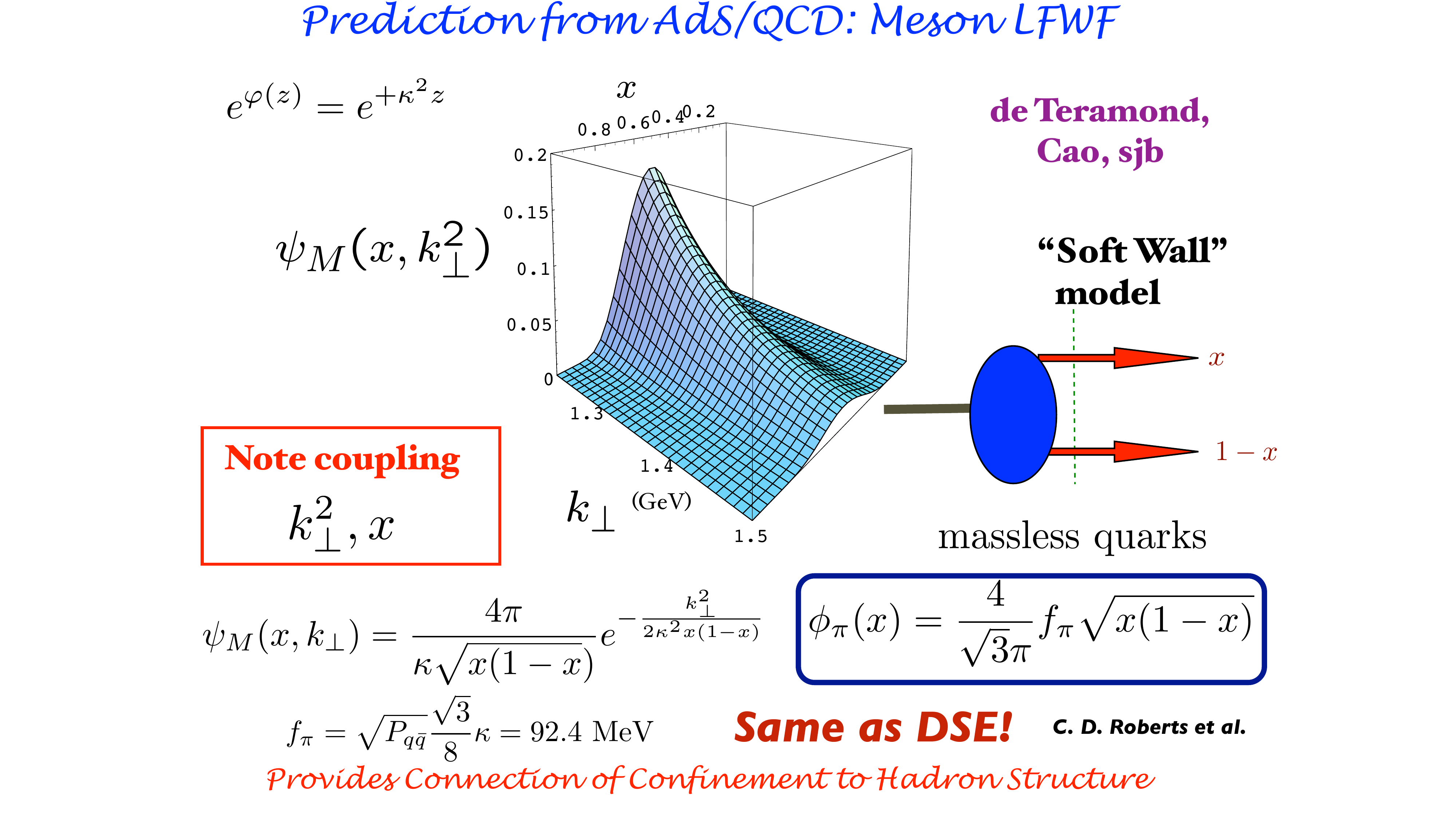}
\end{center}
\caption{The LF wave function of the pion predicted by LF holography.  The results are consistent with analyses based on the Dyson-Schwinger equation.}
\label{Bledslides6.pdf}
\end{figure}

The holographic theory incorporates the dependence on the total quark spin, $S = 0$ for the $\pi$ Regge trajectory, and $S = 1$ for the $\rho$  trajectory, as given by the additional term $2 \kappa^2 S$,  where $S = 0, 1$, in the LF Hamiltonian.  This leads, for example to the correct prediction for the  $\pi - \rho$ mass gap: $M_\rho^2 - M^2_\pi = 2\kappa^2$. In order to describe the quark spin-spin interaction, which distinguishes for example the nucleons from $\Delta$ particles, one includes an identical term, $2 \kappa^2 S$, with $S = 0, 1$ in the LF baryon Hamiltonian which maintains hadronic supersymmetry. The prediction for the mass spectrum of mesons, baryons and tetraquarks is given by~\cite{Brodsky:2016yod}
\begin{align}
M^2_{M \perp} &= 4 \kappa^2 (n+ L_M ) + 2 \kappa^2 S,\\
M^2_{B \perp} &= 4  \kappa^2 (n+ L_B+1) + 2 \kappa^2  S,\\
M^2_{T \perp} & = 4  \kappa^2  (n+ L_T+1) + 2 \kappa^2 S,
\end{align}
with the same slope $\lambda= \kappa^2$ in $L$ and $n$, the radial quantum number.  
The  Regge spectra of the pseudoscalar $S=0$  and vector $S=1$  mesons  are then
predicted correctly, with equal slope in the principal quantum number $n$ and the internal orbital angular momentum.  The nonperturbative pion distribution amplitude 
$\phi_\pi(x) \propto f_\pi \sqrt{x(1-x)}$ predicted by LF holography is  consistent with the Belle data for the photon-to-pion transition form factor~\cite{Brodsky:2011xx}. 
The prediction for the LF wave function $\psi_\rho(x,k_\perp)$ of the  $\rho$ meson gives excellent 
predictions for the observed features of diffractive $\rho$ electroproduction $\gamma^* p \to \rho  p^\prime$~\cite{Forshaw:2012im}.
The prediction for the valence LF wave function of the pion is shown in Fig. \ref{Bledslides6.pdf}.

These predictions for the meson, baryon and tetraquark spectroscopy are specific to zero mass quarks.
In our paper~\cite{deTeramond:2021yyi}, we have shown that the breaking of chiral symmetry in holographic light-front QCD from nonzero quark masses is encoded in the longitudinal dynamics, independent of $\zeta$.
The results for $M^2= M^2_\perp + M^2_L$, where $M^2_L$ is the longitudinal contribution from the nonzero quark mass, retains the zero-mass chiral property of the pion predicted by the superconformal algebraic structure which governs its transverse dynamics. The mass scale in the longitudinal light-front Hamiltonian determines the confinement strength in this direction; It is also responsible for most of the light meson ground state mass, consistent with the standard Gell-Mann-Oakes-Renner constraint.    Longitudinal confinement and the breaking of chiral symmetry are found to be different manifestations of the same underlying dynamics that appears in the
 't Hooft large-$N_C$ QCD(1 + 1) model.  One also obtains spherical symmetry of the 3-dimensional confinement potential in the nonrelativistic limit.  For related work, see Refs. \cite{Li:2021jqb, Ahmady:2021lsh, Ahmady:2021yzh, Weller:2021wog}.

\begin{figure}
 \begin{center}
\includegraphics[height= 8cm,width=15cm]{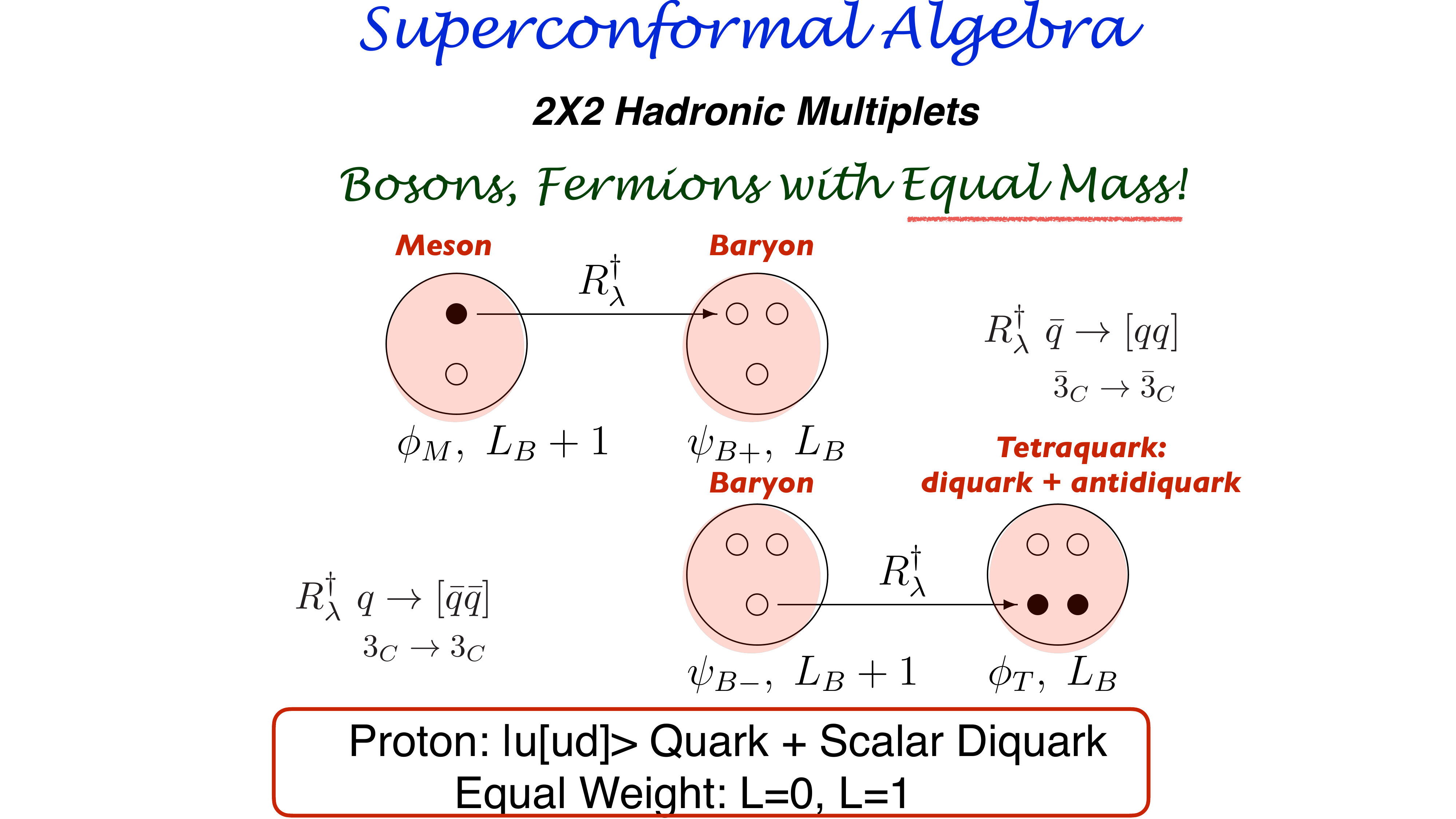}
\end{center}
\caption{The supersymmetric meson-baryon-tetraquark 4-plet.  The operator $R^\dagger_\lambda$  transforms an antiquark $\bar 3_C$ into a diquark $\bar 3_C$.}
\label{Bledslides7.pdf}
\end{figure}

\begin{figure}
 \begin{center}
\includegraphics[height= 10cm,width=15cm]{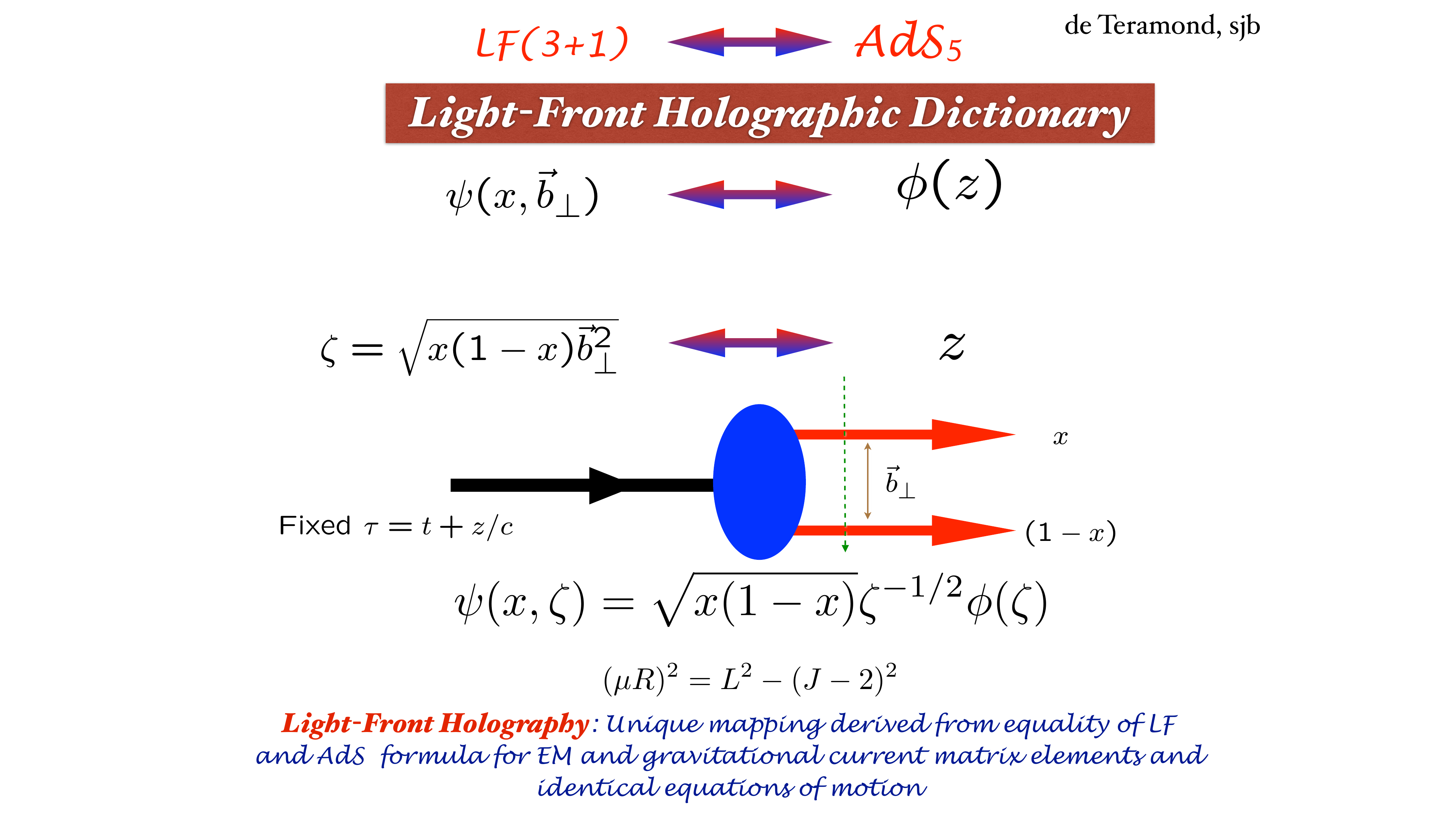}
\end{center}
\caption{The holographic duality connecting LF physics in 3+1 physical space-time with antI-DeSitter space in 5 dimensions. 
The coordinate $z$ in  the fifth dimension of AdS space  is holographically dual to the LF radial variable $\zeta$ where $\zeta^2 = b^2_\perp x(1-x)$. }
\label{Bledslides4.pdf}
\end{figure}

Phenomenological extensions of the holographic QCD approach  have also led to nontrivial connections between the dynamics of form factors and polarized and unpolarized quark distributions with pre-QCD nonperturbative approaches such as Regge theory and the Veneziano model~\cite{Sufian:2016hwn, deTeramond:2018ecg, Liu:2019vsn}. As discussed in the next section, it also predicts the analytic behavior of the QCD coupling $\alpha_s(Q^2)$ in the nonperturbative domain~\cite{Brodsky:2010ur, Deur:2014qfa}.

\begin{figure}
 \begin{center}
\includegraphics[height=10cm,width=15cm]{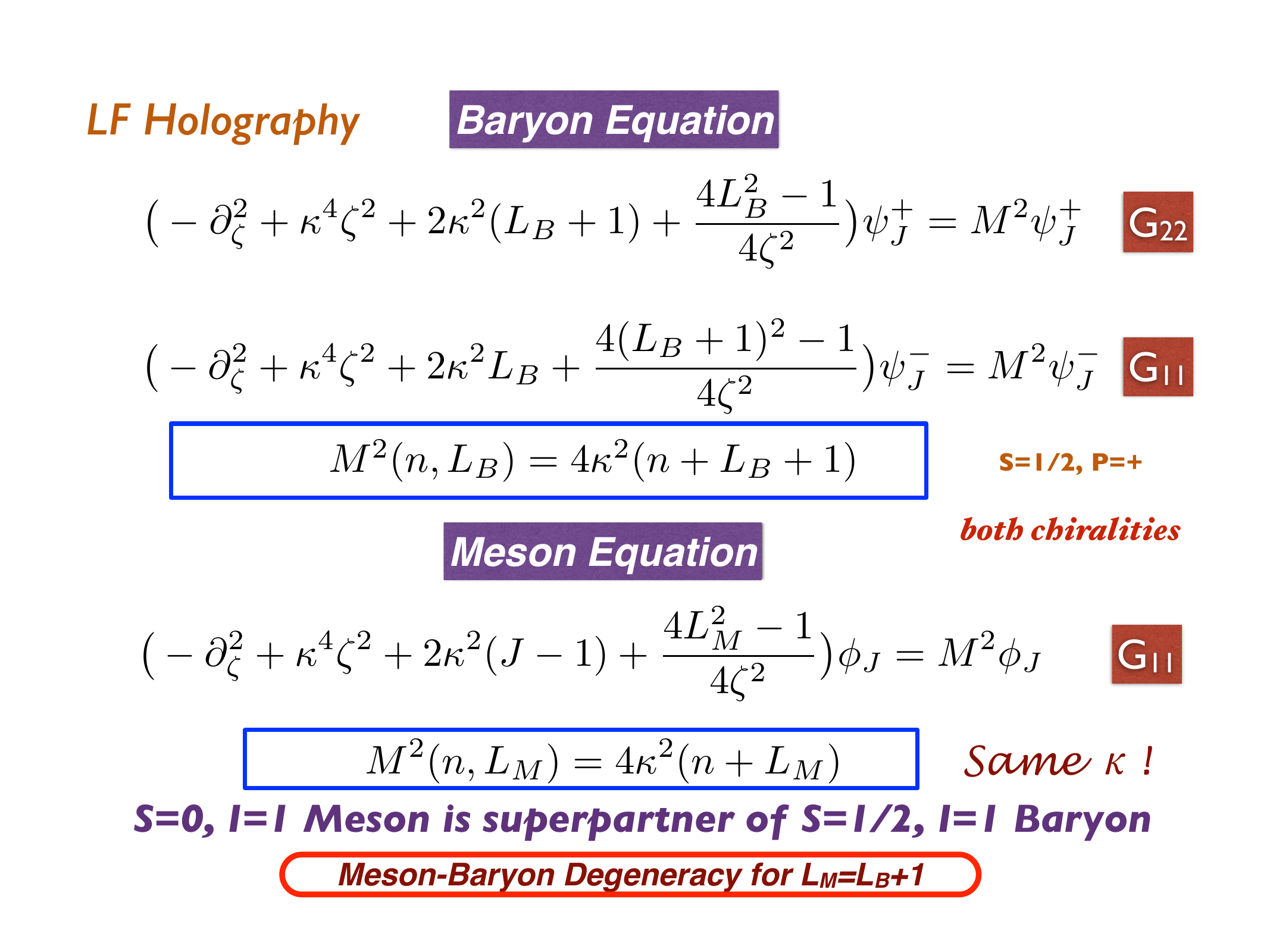}

\end{center}
\caption{(A). The LF Schr\"odinger equations for baryons and mesons for zero quark mass derived from the Pauli $2\times 2$ matrix representation of superconformal algebra.  
The $\psi^\pm$  are the baryon quark-diquark LFWFs where the quark spin $S^z_q=\pm 1/2$ is parallel or antiparallel to the baryon spin $J^z=\pm 1/2$.   The meson and baryon equations are identical if one identifies a meson with internal orbital angular momentum $L_M$ with its superpartner baryon with $L_B = L_M-1.$
See Ref.~\cite{deTeramond:2014asa,Dosch:2015nwa,Dosch:2015bca}.}
\label{FigsJlabProcFig3.pdf}
\end{figure} 
The LF Schr\"odinger Equations for baryons and mesons derived from superconformal algebra  are shown  in Fig. \ref{FigsJlabProcFig3.pdf}.
The comparison between the meson and baryon masses of the $\rho/\omega$ Regge trajectory with the spin-$3/2$ $\Delta$ trajectory is shown in Fig. \ref{FigsJlabProcFig3.pdf}.
Superconformal algebra  predicts the meson and baryon masses are identical if one identifies a meson with internal orbital angular momentum $L_M$ with its superpartner baryon with $L_B = L_M-1.$   Notice that the twist  $\tau = 2+ L_M = 3 + L_B$ of the interpolating operators for the meson and baryon superpartners are the same.   Superconformal algebra also predicts that the LFWFs of the superpartners are identical, and thus they have identical dynamics, such their elastic and transition form factors.   These features can be tested for spacelike  form factors at  JLab12.

The extension of light-front QCD  to superconformal algebra has leads to a specific mass degeneracy between mesons, baryons and tetraquarks~\cite{deTeramond:2014asa, Dosch:2015nwa, Brodsky:2016yod}  underlying the $SU(3)_C$ representation properties, since a diquark cluster has the same color-triplet representation as an antiquark, namely $\bar 3 \in 3 \times 3$.  The meson wave function $\phi_M$, the upper and lower components of the baryon wave function, $\phi_{B \,\pm}$, and the tetraquark wave function, $\phi_T$, can be arranged as a supersymmetric 4-plet matrix~\cite{Brodsky:2016yod, Zou:2018eam}  
\begin{align}
\vert \Phi \rangle =   \begin{pmatrix}
    \phi_M^{\, (L+1)} & \phi_{B \, -}^{\, (L + 1)}\\
    \phi_{B \, + } ^{\, (L)} & \phi_T^{\, (L)}
    \end{pmatrix} ,
\end{align}
with $H^{LF} \vert \Phi \rangle = M^2 \vert \Phi \rangle$ and $L_M = L_B + 1$,  $L_T = L_B$. The constraints from superconformal structure uniquely determine the form of the effective transverse confining potential for mesons, nucleons and tetraquarks~\cite{deTeramond:2014asa, Dosch:2015nwa, Brodsky:2016yod}, 
and lead to the remarkable relations  $L_M = L_B + 1$, $L_T = L_B$. The superconformal algebra also predicts the universality of Regge slopes 
with a unique scale $\lambda=\kappa^2$ for all hadron families.

As noted above, an important feature of LF holography is the application~~\cite{deTeramond:2014asa,Dosch:2015nwa,Brodsky:2016rvj} of {\it superconformal algebra}, a feature of the underlying conformal symmetry of chiral QCD. 
The conformal group has an elegant $ 2\times 2$ Pauli matrix representation called {\it superconformal algebra}, 
originally discovered by  Haag, Lopuszanski, and Sohnius ~\cite{Haag:1974qh}.
The conformal Hamiltonian operator and the special conformal operators can be represented as anticommutators of Pauli matrices
 $H = {1/2}[Q, Q^\dagger]$ and  $K = {1/2}[S, S^\dagger]$.
As shown by Fubini and Rabinovici,~\cite{Fubini:1984hf},  a nonconformal Hamiltonian with a mass scale and universal confinement can then be obtained by shifting $Q \to Q +\omega K$, the analog of the dAFF procedure. 
In effect,  one has obtained generalized supercharges of the superconformal algebra~\cite{Fubini:1984hf}.
This ansatz extends the predictions for the hadron spectrum to a ``4-plet" -- consisting of mass-degenerate quark-antiquark mesons, quark-diquark baryons, and diquark-antidiquark tetraquarks, as shown in fig.~\ref{Bledslides7.pdf}.   The 4-plet contains two entries $\Psi^\pm$  for each baryon, corresponding to internal orbital angular momentum $L$ and $L+1$.  This property of the baryon LFWFs is the analog of the eigensolution of the Dirac-Coulomb equation which has both an upper component $\Psi^+$ and a  lower component $\Psi^- =  {\vec \sigma \cdot \vec p  \over m+E -V} \Psi^+$.  

LF Schr\"odinger Equations for  both baryons and mesons can be derived from superconformal algebra~\cite{deTeramond:2014asa,Dosch:2015nwa,Brodsky:2016rvj,Brodsky:2015oia}.
The baryonic eigensolutions correspond to bound states of $3_C$ quarks to a $\bar 3_C$ spin-0 or spin-1 $qq$ diquark cluster;  the tetraquarks in the 4-plet are bound states of diquarks and  anti-diquarks.  
The quark-diquark baryons have two amplitudes $L_B, L_B+1$  with equal probability, a  feature of ``quark chirality invariance". 
The proton Fock state component $\psi^+$ (with parallel quark and baryon spins) and $\psi^-$ (with anti-parallel quark and baryon spins)  have equal Fock state probability -- a  feature of ``quark chirality invariance".  Thus the proton's spin is carried by quark orbital angular momentum in the nonperturbative domain. Predictions for the static properties of the nucleons are discussed in Ref.~\cite{Liu:2015jna}.  
The overlap of the $L=0$  and $ L=1 $  LF wavefunctions  in the Drell-Yan-West formula is required to have a  non-zero  Pauli form factor $F_2(Q^2)$ and anomalous 
magnetic moment~\cite{Brodsky:1980zm}.  The existence of both components is also necessary to generate the  pseudo-T-odd Sivers single-spin asymmetry in deep inelastic lepton-nucleon scattering~\cite{Brodsky:2002cx}.

The predicted spectra $M^2(n,L) = 4\kappa^2(n+L)$ for mesons, and $M^2(n,L) = 4\kappa^2(n+L+1)$ for baryons, is remarkably consistent with observed hadronic spectroscopy.  
The Regge-slopes in $n$ and $L$ are identical.    
The predicted  meson, baryon and tetraquark masses  coincide if one identifies a meson with internal orbital angular momentum $L_M$ with its superpartner baryon or tetraquark with $L_B = L_M-1$. 
Superconformal algebra thus predicts that mesons with $L_M=L_B+1$ have the same mass as the baryons in the supermultiplet. 
An example of the mass degeneracy of the $\rho/\omega$ meson Regge trajectory with the $J=3/2$ $\Delta$-baryon trajectory is shown in  
Fig.~\ref{FigsJlabProcFig4.pdf}.   The value of $\kappa $ can be set by the $\rho$ mass;  only ratios of masses are predicted.

The combination of light-front holography with superconformal algebra thus leads to the novel prediction that hadron physics has supersymmetric properties in both spectroscopy and dynamics. The excitation spectra of relativistic light-quark meson, baryon and tetraquark bound states all lie on linear Regge
trajectories with identical slopes in the radial and orbital quantum numbers.  Detailed predictions for the tetraquark spectroscopy and  comparisons with the observed hadron spectrum are presented in ref.~\cite{Nielsen:2018uyn}.

\section {The QCD Coupling at All Scales} 
The QCD running coupling can be defined~\cite{Grunberg:1980ja} at all momentum scales from any perturbatively calculable observable, such as the coupling $\alpha^s_{g_1}(Q^2)$ which is defined from measurements of the Bjorken sum rule.   At high momentum transfer, such ``effective charges"  satisfy asymptotic freedom, obey the usual pQCD renormalization group equations, and can be related to each other without scale ambiguity 
by commensurate scale relations~\cite{Brodsky:1994eh}.  
The dilaton  $e^{+\kappa^2 z^2}$ soft-wall modification~\cite{Karch:2006pv} of the AdS$_5$ metric, together with LF holography, predicts the functional behavior 
in the small $Q^2$ domain~\cite{Brodsky:2010ur}: 
${\alpha^s_{g_1}(Q^2) = 
\pi   e^{- Q^2 /4 \kappa^2 }}. $ 
Measurements of  $\alpha^s_{g_1}(Q^2)$ are remarkably consistent with this predicted Gaussian form. 
The predicted coupling is thus finite at $Q^2=0$.

\begin{figure}
\begin{center}
\includegraphics[height=7cm,width=12cm]{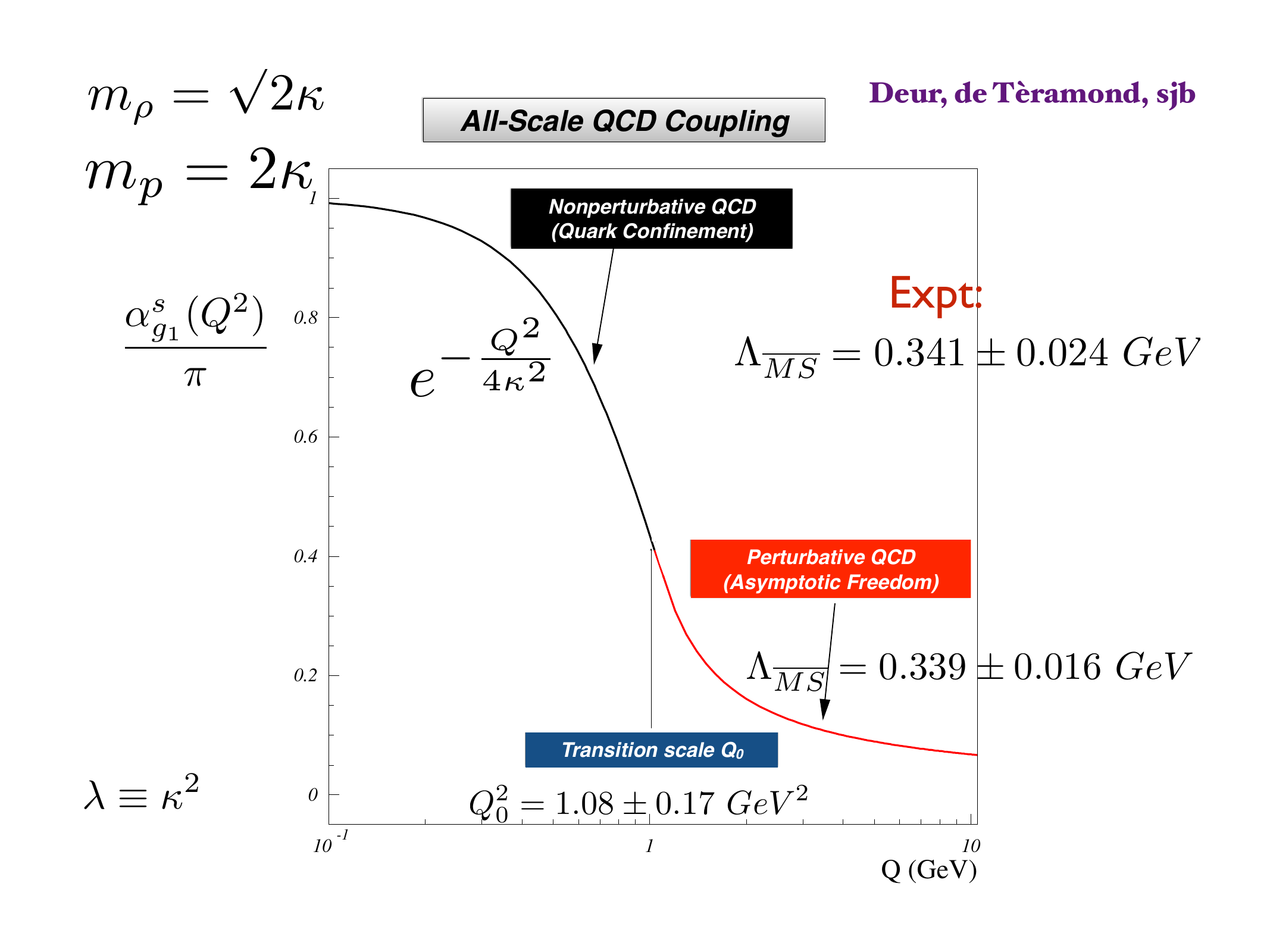}
\includegraphics[height=7cm,width=12cm]{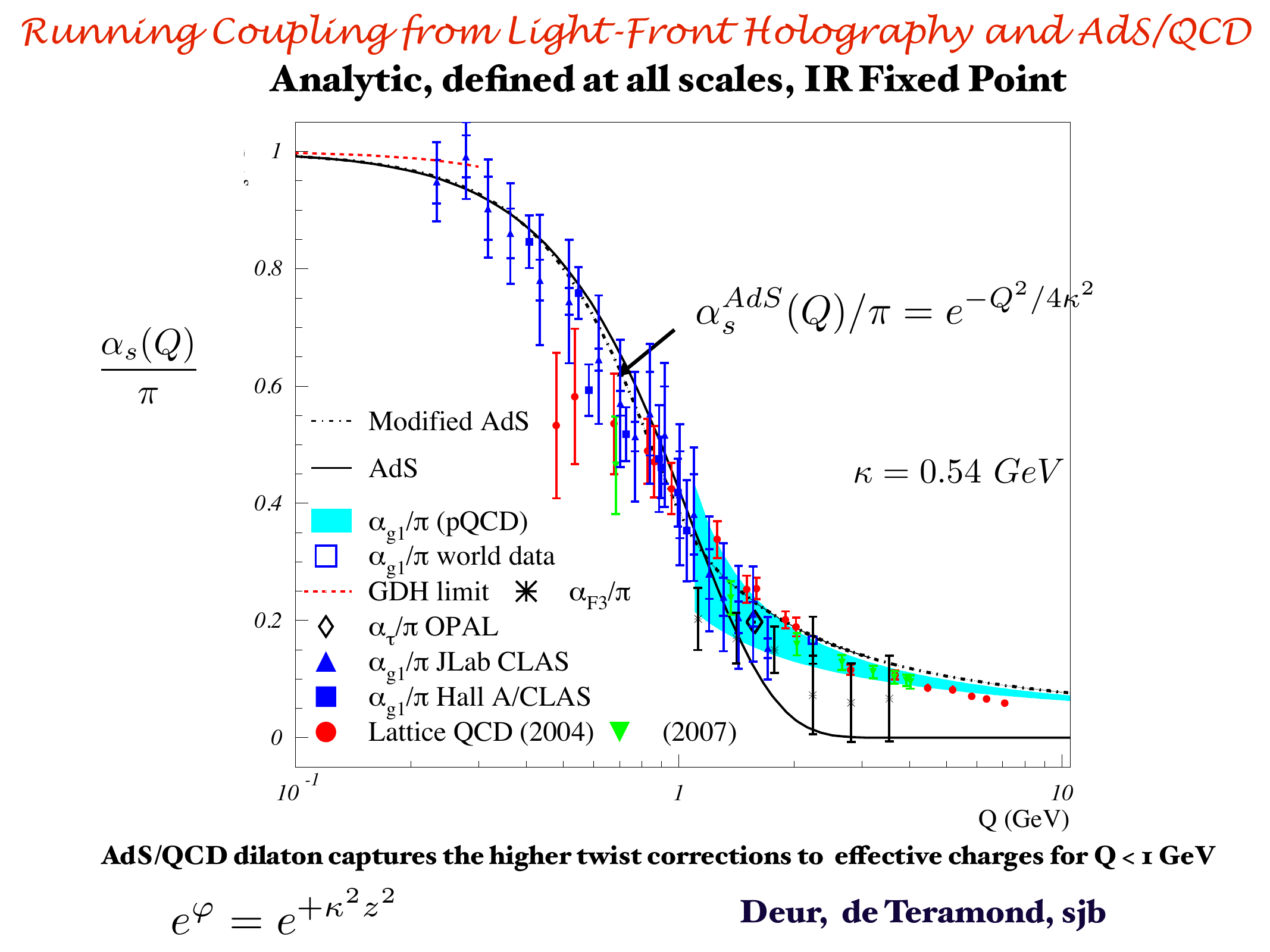}
\end{center}
\caption{
(A). Prediction from LF Holography for the QCD running coupling $\alpha^s_{g_1}(Q^2)$.   The magnitude and derivative of the perturbative and nonperturbative coupling are matched at the scale $Q_0$.  This matching connects the perturbative scale 
$\Lambda_{\overline{MS}}$ to the nonperturbative scale $\kappa$ which underlies the hadron mass scale. 
(B). Comparison of the predicted nonperturbative coupling with measurements of the effective charge $\alpha^s_{g_1}(Q^2)$  
defined from the Bjorken sum rule.  
See Ref.~\cite{Brodsky:2014jia}. 
}
\label{FigsJlabProcFig5.pdf}
\end{figure}

The parameter $\kappa$,  which   determines the mass scale of  hadrons in the chiral limit, can be connected to the  mass scale $\Lambda_s$  controlling the evolution of the perturbative QCD coupling~\cite{Brodsky:2010ur,Deur:2014qfa,Brodsky:2014jia}. This connection can be done for any choice of renormalization scheme, including the $\overline{MS}$ scheme,
as seen in  Fig.~\ref{FigsJlabProcFig5.pdf}. 
The relation between scales is obtained by matching at a scale $Q^2_0$ the nonperturbative behavior of the effective QCD coupling, as determined from light-front holography, to the perturbative QCD coupling with asymptotic freedom.
The result of this perturbative/nonperturbative matching at the analytic inflection point  is an effective QCD coupling  which is defined at all momenta.

Recently \cite{deTeramond:2024ikl}, Guy de Teramond,  Guenter Dosch, Alexandre Deur, Arpon Paul, Tianbo Liu, Raza Sabbir Sufian and I have used analytic continuation to extend the gauge/gravity duality nonperturbative
description of the strong force coupling into the transition, near-perturbative, regime
where perturbative effects become important. By excluding the unphysical region
in coupling space from the flow of singularities in the complex plane, we have derived
a specific relation between the scales relevant at large and short distances; this
relation is uniquely fixed by requiring maximal analyticity. The unified effective
coupling model gives an accurate description of the data in the nonperturbative and
the near-perturbative regions. The analytic determination of $\alpha_s(Q^2)$ over all domains increases the precision and reliability of QCD predictions.

\section{Superconformal Algebra and Supersymmetric Hadron Spectroscopy }

Another advance in LF holography is the application~~\cite{deTeramond:2014asa,Dosch:2015nwa,Brodsky:2016rvj} of {\it superconformal algebra}, a feature of the underlying conformal symmetry of chiral QCD. 
The conformal group has an elegant $ 2\times 2$ Pauli matrix representation called {\it superconformal algebra}, 
originally discovered by  Haag, Lopuszanski, and Sohnius ~\cite{Haag:1974qh}.
The conformal Hamiltonian operator and the special conformal operators can be represented as anticommutators of Pauli matrices
 $H = {1/2}[Q, Q^\dagger]$ and  $K = {1/2}[S, S^\dagger]$.
As shown by Fubini and Rabinovici,~\cite{Fubini:1984hf},  a nonconformal Hamiltonian with a mass scale and universal confinement can then be obtained by shifting $Q \to Q +\omega K$, the analog of the dAFF procedure. 
In effect,  one has obtained generalized supercharges of the superconformal algebra~\cite{Fubini:1984hf}.
This ansatz extends the predictions for the hadron spectrum to a ``4-plet" -- consisting of mass-degenerate quark-antiquark mesons, quark-diquark baryons, and diquark-antidiquark tetraquarks, as shown in fig.~\ref{Bledslides7.pdf}.   The 4-plet contains two entries $\Psi^\pm$  for each baryon, corresponding to internal orbital angular momentum $L$ and $L+1$.  This property of the baryon LFWFs is the analog of the eigensolution of the Dirac-Coulomb equation which has both an upper component $\Psi^+$ and a  lower component $\Psi^- =  {\vec \sigma \cdot \vec p  \over m+E -V} \Psi^+$.  

LF Schr\"odinger Equations for  both baryons and mesons can be derived from superconformal algebra~\cite{deTeramond:2014asa,Dosch:2015nwa,Brodsky:2016rvj,Brodsky:2015oia}.
The baryonic eigensolutions correspond to bound states of $3_C$ quarks to a $\bar 3_C$ spin-0 or spin-1 $qq$ diquark cluster;  the tetraquarks in the 4-plet are bound states of diquarks and  anti-diquarks.  
The quark-diquark baryons have two amplitudes $L_B, L_B+1$  with equal probability, a  feature of ``quark chirality invariance". 
The proton Fock state component $\psi^+$ (with parallel quark and baryon spins) and $\psi^-$ (with anti-parallel quark and baryon spins)  have equal Fock state probability -- a  feature of ``quark chirality invariance".  Thus the proton's spin is carried by quark orbital angular momentum in the nonperturbative domain. Predictions for the static properties of the nucleons are discussed in Ref.~\cite{Liu:2015jna}.  
The overlap of the $L=0$  and $ L=1 $  LF wavefunctions  in the Drell-Yan-West formula is required to have a  non-zero  Pauli form factor $F_2(Q^2)$ and anomalous 
magnetic moment~\cite{Brodsky:1980zm}.  The existence of both components is also necessary to generate the  pseudo-T-odd Sivers single-spin asymmetry in deep inelastic lepton-nucleon scattering~\cite{Brodsky:2002cx}.

The predicted spectra $M^2(n,L) = 4\kappa^2(n+L)$ for mesons, and $M^2(n,L) = 4\kappa^2(n+L+1)$ for baryons, is remarkably consistent with observed hadronic spectroscopy.  
The Regge-slopes in $n$ and $L$ are identical.    
The predicted  meson, baryon and tetraquark masses  coincide if one identifies a meson with internal orbital angular momentum $L_M$ with its superpartner baryon or tetraquark with $L_B = L_M-1$. 
Superconformal algebra thus predicts that mesons with $L_M=L_B+1$ have the same mass as the baryons in the supermultiplet. 
An example of the mass degeneracy of the $\rho/\omega$ meson Regge trajectory with the $J=3/2$ $\Delta$-baryon trajectory is shown in  
Fig.~\ref{FigsJlabProcFig4.pdf}.   The value of $\kappa $ can be set by the $\rho$ mass;  only ratios of masses are predicted.

The combination of light-front holography with superconformal algebra thus leads to the novel prediction that hadron physics has supersymmetric properties in both spectroscopy and dynamics. The excitation spectra of relativistic light-quark meson, baryon and tetraquark bound states all lie on linear Regge
trajectories with identical slopes in the radial and orbital quantum numbers.  Detailed predictions for the tetraquark spectroscopy and  comparisons with the observed hadron spectrum are presented in ref.~\cite{Nielsen:2018uyn}.

\section{Renormalization Scale Setting}

A key problem in making precise perturbative QCD predictions is
the uncertainty in determining the renormalization scale $\mu$ of
the running coupling $\alpha_s(\mu^2).$ 
The purpose of the running
coupling in any gauge theory is to sum all terms involving the
$\beta$ function; in fact, when the renormalization scale is set
properly, all non-conformal $\beta \ne 0$ terms  in a perturbative
expansion arising from renormalization are summed into the running
coupling. The remaining terms in the perturbative series are then
identical to that of a  conformal theory; i.e., the corresponding
theory with $\beta=0$. 
There is no renormalization scale-setting ambiguity for precision tests of quantum electrodynamics.   The scale of the running QED coupling is set to absorb all vacuum polarization  diagrams; i.e. the $\beta$ terms. The coefficients in the  perturbative QCD series then matches conformal theory; i.e. the corresponding perturbative series with $\beta=0$.   This is the standard Gell-Mann Low scale-setting procedure for high precision tests of QED, where all vacuum polarization contributions are summed into the QED running coupling. The same scale-setting procedure applies to the $SU(2)_{EW}$ theory of the electroweak interactions.
An important analytic property of non-Abelian QCD with $N_C$ colors is that it must agree analytically with Abelian QED in the $N_C\to 0$ limit, at fixed 
$\hat \alpha_s = C_F \alpha_s$ and fixed $\hat n_f = T {n_f\over C_F}$ with $C_F = {N^2_C -1\over 2 N_C}$ and $T =1/2$. This is the ``Abelian correspondence principle."  Thus the setting of the renormalization scale in QCD must agree with Gell-Mann-Low scale setting for QED in the  $N_C \to 0$ limit.

It has become conventional to simply guess the renormalization scale and choose an arbitrary range of uncertainty when making perturbative QCD (pQCD) predictions. However, this {\it ad hoc} assignment of the renormalization scale and the estimate of the size of the resulting uncertainty leads to anomalous renormalization scheme-and-scale dependences. In fact, relations between physical observables must be independent of the theorist's choice of the renormalization scheme, and the renormalization scale in any given scheme at any given order of pQCD is not ambiguous. This was the motivation for the BLM (Brodsky-Lepage-Mackenzie)~\cite{Brodsky:1982gc}  procedure for  QCD scale-setting.  It was then generalized to all orders as the PMC (the Principle of Maximum Conformality.
The {\it Principle of Maximum Conformality} (PMC)~\cite{Brodsky:2011ig,Brodsky:2013vpa}, which generalizes the conventional Gell-Mann-Low method for scale-setting in perturbative QED to non-Abelian QCD, provides a rigorous method for achieving unambiguous scheme-independent, fixed-order predictions for observables consistent with the principles of the renormalization group. 

The PMC scale-setting procedure
sets the renormalization scale $\alpha_s(Q^2_{PMC})$ at every order by absorbing the $\beta$ terms appearing in the pQCD series. The resulting pQCD series thus matches the corresponding conformal series with all $\beta$ terms set to $0$.   The  problematic $n!$ ``renormalon" divergence of pQCD series associated with the nonconformal terms does not appear in the conformal series and the conformal series is independent of the theorist's choice of renormalization scheme.  
This also means that relations between any two perturbatively calculable observables are scheme-independent.  These relations are called ``commensurate scale relations"~\cite{Brodsky:2013vpa, Brodsky:1994eh}.  The PMC also satisfies the requirement that one must use the same scale-setting procedure in all sectors of a Grand-Unified Theory of QED, the electroweak interactions, and QCD~\cite{Binger:2003by}.   The renormalization scale of the running coupling depends dynamically
on the virtuality of the underlying quark and gluon subprocess and thus the specific kinematics of each event.

The resulting scale-fixed predictions for physical observables using
the PMC are also {\it  independent of
the choice of renormalization scheme} --  a key requirement of 
renormalization group invariance.  The PMC predictions are also independent of the choice of the {\it initial} renormalization scale $\mu_0.$  
The PMC sums all of the non-conformal terms associated with the QCD $\beta$ function, thus providing a rigorous method for eliminating renormalization scale ambiguities in quantum field theory. 
We have also showed that a single global
PMC scale, valid at leading order, can be derived from basic
properties of the perturbative QCD cross section.   We have given a detailed comparison of these PMC approaches by comparing their predictions for three important quantities 
$R_{e+ e}$, $R_\tau$ and $\Gamma_{H \to b \bar b}$ up to four-loop pQCD corrections~\cite{Brodsky:2011ig}.  The numerical results show that the single-scale PMCs method, which involves a somewhat simpler analysis, can serve as a reliable substitute for the full multi-scale PMCm method, and that it leads to more precise pQCD predictions with less residual scale dependence.
The PMC thus greatly improves the reliability and precision of QCD predictions at the LHC and other colliders~\cite{Brodsky:2011ig}.
 As we have demonstrated, the PMC also has the potential to greatly increase the sensitivity of experiments at the LHC to new physics beyond the Standard Model.

Predictions based on PMC scale setting  satisfies the self-consistency conditions of the renormalization group, including reflectivity, symmetry and transitivity~\cite{Brodsky:2012ms}. The resulting PMC predictions  satisfy all of the basic requirements of RGI.

 The transition scale between the perturbative and nonperturbative domains can also be determined by using the PMC~\cite{Deur:2014qfa, Deur:2016cxb, Deur:2016tte, Deur:2017cvd}, thus providing a procedure for setting the ``factorization" scale for pQCD evolution. The running coupling resums all of the $\{\beta_i\}$-terms by using the PMC, which naturally leads to a more convergent and renormalon-free pQCD series.

In more detail: the PMC scales are determined by applying the RGE of the QCD running coupling.   By recursively applying the RGE  one establishes a perturbative $\beta$-pattern at each order in a pQCD expansion. For example, the usual scale-displacement relation for the running couplings at two different scales $Q_1$ and $Q_2$ can be deduced from the RGE, which reads
\begin{eqnarray}
a_{Q_2} &=& a_{Q_1}- \beta_{0} \ln\left(\frac{Q_2^{2}} {Q_1^2}\right) a_{Q_1}^2 +\left[\beta^2_{0} \ln^2 \left(\frac{Q_2^{2}}{Q_1^2}\right) -\beta_{1} \ln \left(\frac{Q_2^{2}} {Q_1^2}\right)\right] a_{Q_1}^3 \nonumber\\
&& + \left[- \beta_0^3\ln^3 \left(\frac{Q_2^{2}}{Q_1^2}\right) + \frac{5}{2}{\beta_0}{\beta_1} \ln^2 \left(\frac{Q_2^{2}}{Q_1^2}\right) - \beta_2 \ln \left(\frac{Q_2^{2}}{Q_1^2}\right) \right]a_{Q_1}^4 + \left[\beta_0^4 \ln^4 \left(\frac{Q_2^{2}}{Q_1^2}\right) \right. \nonumber\\
&& \left. - \frac{13}{3}{\beta_0^2}{\beta_1} \ln^3 \left(\frac{Q_2^{2}}{Q_1^2}\right) + \frac{3}{2}{\beta_1^2} \ln^2 \left(\frac{Q_2^{2}}{Q_1^2}\right) + 3{\beta_2}{\beta_0} \ln^2\left(\frac{Q_2^{2}}{Q_1^2}\right) - \beta_3 \ln \left(\frac{Q_2^{2}}{Q_1^2}\right) \right]a_{Q_1}^5 + \cdots, \label{scaledis}
\end{eqnarray}
where $a_{Q_i}= \alpha_s(Q_i)/\pi$, the functions $\beta_0, \beta_1, \cdots$ are generally scheme dependent, which correspond to the one-loop, two-loop, $\cdots$, contributions to the RGE, respectively.  The PMC utilizes this perturbative $\beta$-pattern to systematically set the scale of the running coupling at each order in a pQCD expansion. 

The coefficients of the $\{\beta_i\}$-terms in the $\beta$-pattern can  be identified by reconstructing  ``degeneracy relations"~\cite{Mojaza:2012mf, Brodsky:2013vpa} among different orders. The degeneracy relations, which underly the conformal features of the resultant pQCD series by applying the PMC, are general properties of a non-Abelian gauge theory~\cite{Bi:2015wea}. The PMC prediction achieved in this way resembles a skeleton-like expansion~\cite{Lu:1991yu, Lu:1991qr}. The resulting PMC scales reflect the virtuality of the amplitudes relevant to each order, which are physical in the sense that they reflect the virtuality of the gluon propagators at a given order, as well as setting the effective number ($n_f$) of active quark flavors. The momentum flow for the process involving three-gluon vertex can be determined by properly dividing the total amplitude into gauge-invariant amplitudes~\cite{Binger:2006sj}. Specific values for the PMC scales are computed as a perturbative expansion, so they have small uncertainties which can vary order-by-order.  The PMC scales and the resulting fixed-order PMC predictions are to high accuracy independent of the initial choice of renormalization scale, e.g. the residual uncertainties due to unknown higher-order terms are negligibly small because of the combined suppression effect from both the exponential suppression and the $\alpha_s$-suppression~\cite{Mojaza:2012mf, Brodsky:2013vpa}.

When one applies the standard PMC procedures, different scales generally appear at each order; this  is called the PMC multi-scale approach which often requires considerable theoretical analysis. To make the PMC scale-setting procedure simpler and more easily to be automatized, a single-scale approach (PMC-s), which achieves many of the same PMC goals, has been suggested in Ref.\cite{Shen:2017pdu}. This method effectively replaces the individual PMC scale at each order by a single (effective) scale in the sense of a mean value theorem;  e.g. it can be regarded as a weighted average of the PMC scales at each order derived under PMC multi-scale approach. The single ``PMC-s" scale shows stability and convergence with increasing order in pQCD, as observed by the $e^+e^-$ annihilation cross-section ratio $R_{e^+e^-}$ and the Higgs decay-width $\Gamma(H \to b \bar{b})$, up to four-loop level. Moreover, its predictions are again explicitly independent of the choice of the initial renormalization scale. Thus the PMC-s approach, which involves  a simpler analysis, can be adopted as a reliable substitute for the PMC multi-scale approach, especially when one does not need detailed information at each order.

There are also cases in which additional momentum flows occur, whose scale uncertainties can also be eliminated by applying the PMC. For example, there are two types of log terms, $\ln(\mu/M_{Z})$ and $\ln(\mu/M_{t})$~\cite{Baikov:2008jh, Baikov:2010je, Baikov:2012zn, Baikov:2012er}, for the axial singlet $r^A_S$ of the hadronic $Z$ decays. By applying the PMC, one finds the optimal scale is $Q^{\rm AS} \simeq 100$ GeV~\cite{Wang:2014aqa}, indicating that the typical momentum flow for $r^A_S$ is closer to $M_Z$ than $M_t$. The PMC can also be systematically applied to multi-scale problems. The typical momentum flow can be distinct; thus, one should apply the PMC separately in each region. For example, two optimal scales arise at the N$^2$LO level for the production of massive quark-anti-quark pairs ($Q \bar Q$) close to threshold~\cite{Brodsky:1995ds}, with one being proportional to $\sqrt{\hat{s}}$ and the other to $v\sqrt{\hat{s}}$, where $v$ is the $Q$ and $\bar Q$ relative velocity.

The renormalization scale depends on kinematics such as thrust $(1-T)$ for three jet production via $e^+ e^-$ annihilation. A definitive advantage of using the PMC is that since the PMC scale varies with $(1-T)$,  one can extract directly the strong coupling $\alpha_s$ at a wide range of scales using the experimental data at single center-of-mass-energy, $\sqrt{s}=M_Z$. In the case of conventional scale setting, the predictions are scheme-and-scale dependent and do not agree with the precise experimental results; the extracted coupling constants in general deviate from the world average. In contrast, after applying the PMC, we obtain a comprehensive and self-consistent analysis for the thrust variable results including both the differential distributions and the mean values~\cite{Wang:2019ljl}. Using the ALEPH data~\cite{Heister:2003aj}, the extracted $\alpha_s$ shows that in the scale range of $3.5$ GeV $<Q<16$ GeV (the corresponding ($1-T$) range is $0.05<(1-T)<0.29$), the extracted values for $\alpha_s$ are in excellent agreement with the world average evaluated from $\alpha_s(M_Z)$.

An essential property of renormalizable SU(N)]/U(1) gauge theories, is ``Intrinsic Conformality,"~\cite{DiGiustino:2020fbk}.  It underlies the scale invariance of physical observables and can be used to resolve the conventional renormalization scale ambiguity {\it at every order } in pQCD.  This reflects the underlying conformal properties displayed by pQCD at NNLO, eliminates the scheme dependence of pQCD predictions and is consistent with the general properties of the PMC.   We have also introduced a new method \cite{DiGiustino:2020fbk} to identify the conformal and $\beta$ terms which can be applied either to numerical or to theoretical calculations and in some cases allows infinite resummation of the pQCD series,
The implementation of the PMC$_\infty$ can significantly improve the precision of pQCD predictions; its implementation in multi-loop analysis also simplifies the calculation of higher orders corrections in a general renormalizable gauge theory.
This method has also been used to improve the NLO  pQCD prediction for $t \bar t$ pair production and other processes at the LHC,  where subtle aspects of the renormalization scale of the three-gluon vertex and multi gluon amplitudes, as well as  large radiative corrections to heavy quarks at threshold play a crucial role.  
The large discrepancy of pQCD predictions with  the forward-backward asymmetry measured at the Tevatron is significantly reduced from 3~$\sigma$ to approximately 1~$\sigma.$

The PMC  has also been used to precisely determine the QCD running coupling constant $\alpha_s(Q^2) $  over a wide range of $Q^2$ from event shapes for electron-positron annihilation measured at a single
energy $\sqrt s$ ~\cite{Wang:2019isi}.
The PMC method has  been applied  to a spectrum of LHC processes including Higgs production, jet shape variables, and final states containing a high $p_T$ photon plus heavy quark jets, all of which, sharpen the precision of the Standard Model  predictions. Recently, the PMC has been used  to determine the QCD coupling over the entire range of
validity of perturbative QCD to high precision from the data of a single experiment: 
the thrust  and C-parameter distributions in $e^+ e^-$ annihilation at a
single annihilation energy  $\sqrt s - M^z$~\cite{DiGiustino:2024zss}.
We have also showed that a single global
PMC scale, valid at leading order, can be derived from basic
properties of the perturbative QCD cross section.   We have given a detailed comparison of these PMC approaches by comparing their predictions for three important quantities 
$R_{e+ e}$, $R_\tau$ and $\Gamma_{H \to b \bar b}$ up to four-loop pQCD corrections~\cite{Brodsky:2011ig}.  The numerical results show that the single-scale PMCs method, which involves a somewhat simpler analysis, can serve as a reliable substitute for the full multi-scale PMCm method, and that it leads to more precise pQCD predictions with less residual scale dependence.

The PMC provides first-principle predictions for QCD; it satisfies renormalization group invariance and eliminates the conventional renormalization scheme-and-scale ambiguities, greatly improving the precision of tests of the Standard Model and the sensitivity of collider experiments to new physics. Since the perturbative coefficients obtained using the PMC are identical to those of a conformal theory, one can derive all-orders commensurate scale relations between physical observables evaluated at specific relative scales.
The PMC thus can greatly increase the sensitivity of experiments at the LHC to new physics beyond the Standard Model.

A detailed discussion of how the PMC eliminates renormaiization-scale and scheme ambiguities is given in the  review~\cite{DiGiustino:2023zmw}.  
The QCD running coupling and the definition of effective charges are discussed in the article~\cite{Deur:2023dzc}.

\section{Intrinsic Heavy Quarks}

Quantum Chromodynamics (QCD), the underlying theory of strong interactions, with quarks and gluons as the fundamental degrees of freedom, predicts that the heavy quarks in the nucleon-sea to have both perturbative ``extrinsic" and nonperturbative ``intrinsic" origins.  The extrinsic sea arises from gluon splitting which is triggered by a probe in the reaction. It can be calculated order-by-order in perturbation theory.  In contrast, the intrinsic sea is encoded in the nonperturbative wave functions of the nucleon eigenstate. 

The existence of nonperturbative intrinsic charm (IC) was originally proposed in the BHPS model~\cite{Brodsky:1980pb} and developed further in subsequent papers~\cite{Brodsky:1984nx,Harris:1995jx,Franz:2000ee}. The intrinsic contribution to the heavy quark  distributions of hadrons at high $x$ corresponds to Fock states such as  $|uud Q \bar Q\rangle$ where the heavy quark 
pair is multiply connected to two or more valence quarks of the proton, in distinction to the higher order corrections to DGLAP evolution. The LF wave function is maximal at minimal off-shellness; i.e., when the constituents all have the same rapidity  $y_i$, and thus 
$x_i \propto \sqrt{(m_i^2+ { \vec k_{\perp i}}^2 )}$.  Here $x= {k^+\over P^+} = {k^0 + k^3\over P^0 + P^3}$ is the frame-independent light-front momentum fraction carried by the heavy quark in a hadron with momentum $P^\mu$. 
In the case of deep inelastic lepton-proton scattering, the LF momentum fraction variable $x$  in the proton structure functions can be identified with the Bjorken variable 
$x = {Q^2\over 2 p \cdot q}.$
These heavy quark contributions 
to the nucleon's PDF thus peak at large $x_{bj}$ and thus have important  implication for LHC and EIC collider phenomenology, including Higgs and heavy hadron production at high $x_F$~\cite{Royon:2015eya}.
It also opens up new opportunities to study heavy quark phenomena in fixed target experiments such as the proposed AFTER~\cite{Brodsky:2015fna} fixed target facility at CERN.  Other applications are presented in Refs.~\cite{Brodsky:2020zdq,Bednyakov:2017vck,Brodsky:2016fyh}.
The existence of intrinsic heavy quarks also illuminates fundamental aspects of nonperturbative QCD.

In  Light-Front Hamiltonian theory, the intrinsic heavy quarks of the proton are associated with non-valence Fock states. 
such as $|uud Q \bar Q \rangle$ in the hadronic eigenstate of the LF Hamiltonian; this implies that the heavy quarks are multi-connected to the valence quarks. The probability for the heavy-quark Fock states scales as $1/m^2_Q$ in non-Abelian QCD.  Since the LF wave function is maximal at minimum off-shell invariant mass; i.e., at equal rapidity, the intrinsic heavy quarks carry large momentum fraction $x_Q$.  A key characteristic is different momentum and spin distributions for the intrinsic $Q$ and $\bar Q$ in the nucleon; for example the charm-anticharm asymmetry, since the comoving quarks are sensitive to the global quantum numbers of the nucleon~\cite{Brodsky:2015fna}.  Furthermore, since all of the  intrinsic quarks in the $|uud Q \bar Q\rangle$  Fock state have similar rapidities as the valence quarks, they can re-interact, leading to significant $Q$ vs $\bar Q$ asymmetries.  The concept of intrinsic heavy quarks was also proposed in the context of  meson-baryon fluctuation models~\cite{Pumplin:2005yf, Navarra:1995rq}, where intrinsic charm was identified with two-body state $\bar{D}^0(u\bar{c})\Lambda^+_c(udc)$ in the proton. This identification  predicts large asymmetries in the charm versus anti-charm momentum and spin distributions,  Since these heavy quark distributions depend on the correlations determined by the valence quark distributions, they are referred to as {\it  intrinsic } contributions to the hadron's fundamental structure. A specific analysis of the intrinsic charm content of the deuteron is given in Ref.~\cite{Brodsky:2018zdh}.
In contrast, the contribution to the heavy quark PDFs arising from gluon splitting are symmetric in $Q$ vs $\bar Q$. The contributions generated by DGLAP evolution at low $x$ can be considered as  {\it extrinsic} contributions since they only depend on the gluon distribution. The gluon splitting contribution to the heavy-quark degrees of freedom is  perturbatively calculable 
using  DGLAP
evolution.  To first approximation, the perturbative extrinsic heavy quark distribution falls as $(1-x)$ times the gluon distribution and is limited to low $x_{bj}.$
Thus, unlike the conventional $\log m^2_Q$ dependence of the low $x$  extrinsic gluon-splitting contributions, the probabilities for the intrinsic heavy quark Fock states at high $x$  scale as $1\over m_Q^2$  in non-Abelian QCD, and the relative probability of intrinsic bottom to charm is of order ${m^2_c\over m^2_b} \sim  {1\over 10}.$
In contrast, the probability for a higher Fock state containing heavy leptons in a QED atom  scales as $1\over m_\ell^4$, corresponding to the twist-8 Euler-Heisenberg light-by-light self-energy insertion.  Detailed derivations based on the OPE have been given in Refs.~\cite{Brodsky:1984nx,Franz:2000ee}.

\begin{figure}[htp]
\begin{center}
\setlength\belowcaptionskip{-2pt}
\includegraphics[width=3.2in, height=2.3in]{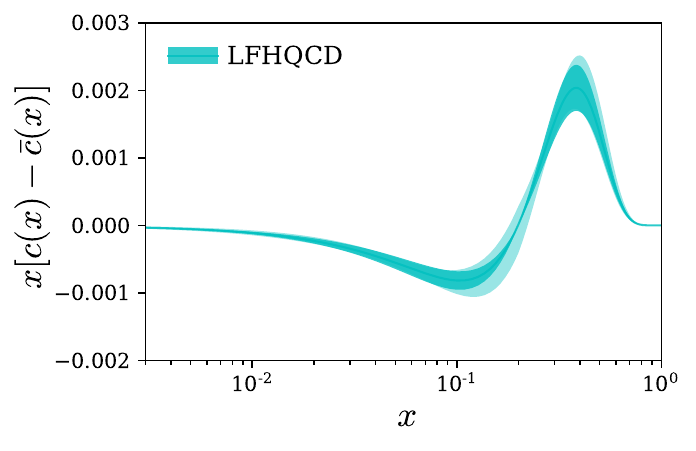}
\caption{The difference of charm and anticharm structure functions $x[c(x)-\bar{c}(x)]$ obtained from the LFHQCD formalism using the lattice QCD input of charm electromagnetic form factors $G^c_{E,M}(Q^2)$ \label{fig:ccbardis}. 
The outer cyan band indicates an estimate of systematic uncertainty in the $x[c(x)-\bar{c}(x)]$ distribution obtained from a variation of  the hadron scale $\kappa_c$ by 5\%.  From Ref.~\cite{Sufian:2020coz}}.
\end{center}
\end{figure}

In an important development~\cite{Sufian:2020coz},  the difference of the charm and anticharm  quark distributions in the proton, $\Delta c(x) = c(x) -\bar c(x)$,  has been computed from first principles in QCD using lattice gauge theory.  A key  theoretical tool is the computation of the charm and anticharm quark contribution 
to the electromagnetic form factor of the proton which would vanish if $c(x) =\bar c(x).$    The exclusive-inclusive connection, together with the LFHQCD formalism, predicts the asymmetry of structure functions $c(x)- \bar c(x)$ which is also odd under charm-anticharm interchange.   
The predicted $c(x)- \bar c(x)$ distribution is large and nonzero at large at $x \sim 0.4$, consistent with the expectations of intrinsic charm. See Fig.~\ref{fig:ccbardis}.   Detailed predictions, including interference contributions with the extrinsic charm contribution is given in ref. \cite{Lykasov:2025uwm}

The $c(x)$ vs. $\bar c(x)$  asymmetry can also be understood physically by identifying the $ |uud c\bar c \rangle$ Fock state with the $|\Lambda_{udc} D_{u\bar c} \rangle$ off-shell excitation of the proton.
A related application of lattice gauge theory to the nonperturbative strange-quark sea from lattice QCD is given in Ref.~\cite{Sufian:2018cpj}.

\begin{figure}
 \begin{center}
\includegraphics[height= 10cm,width=15cm]{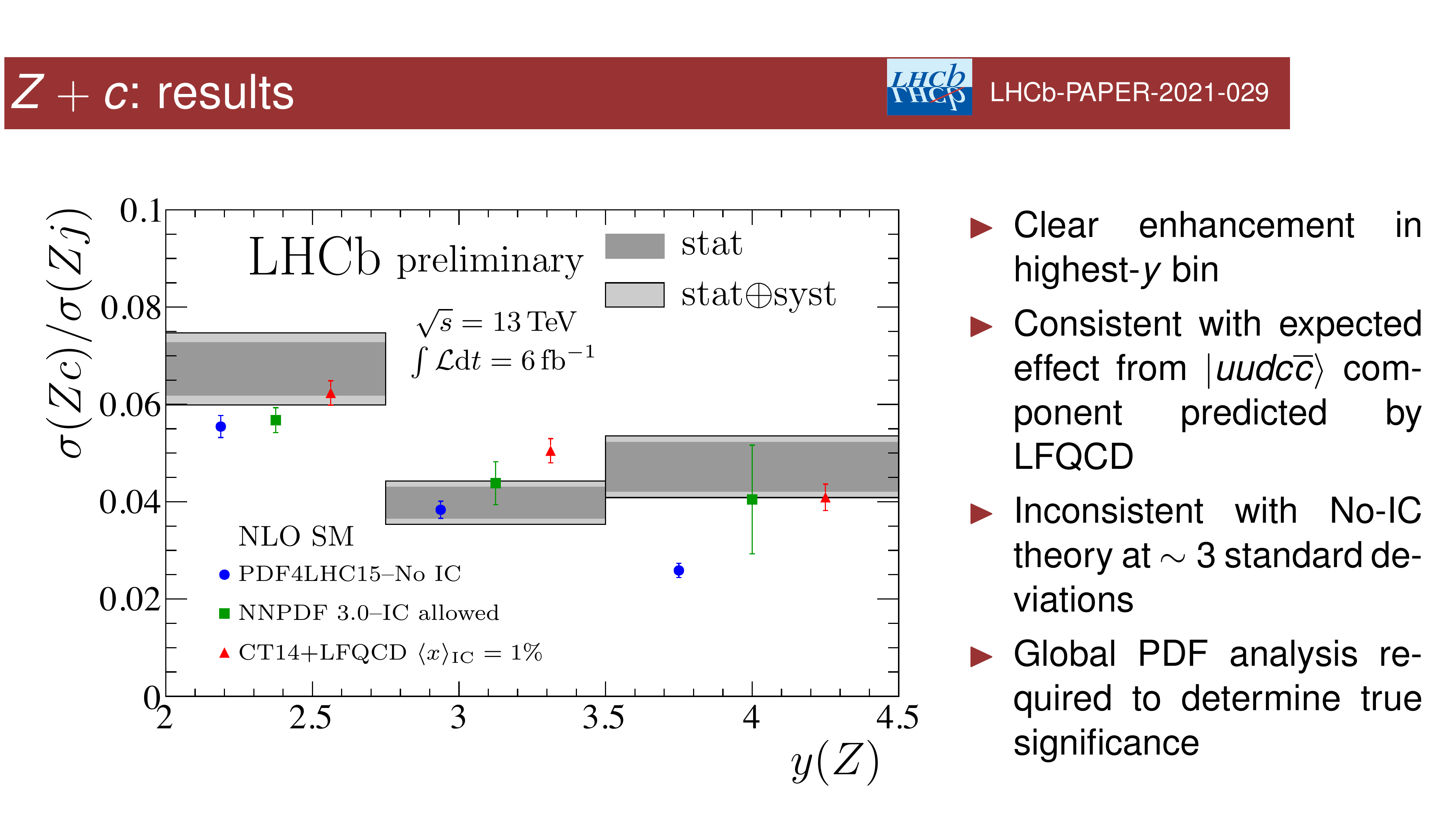}
\end{center}
\caption{The charm distribution in the proton determined from LHCb  measurements of 
$Z$ bosons produced in association with charm at forward rapidity~\cite{LHCb:2021stx}.}
\label{Bledslides1LHCb.pdf}
\end{figure}

There have been many phenomenological calculations involving the existence of a non-zero IC  component which can explain anomalies in the experimental data and to predict  its novel signatures of IC in upcoming experiments~\cite{Brodsky:2015fna}.   A measurement by the LHCb is shown in Fig. 10.
The observed spectrum exhibits a sizable enhancement at forward Z rapidities, consistent with the effect expected if the proton contains the $ |uud \bar c c \rangle$ Fock state predicted by LFQCD.~\cite{LHCb:2021stx}

Thus QCD predicts two separate and distinct contributions to the heavy quark distributions $q(x,Q^2)$ of  the nucleons at low and high $x$.
Here $x= {k^+\over P^+} = {k^0 + k^3\over P^0 + P^3}$ is the frame-independent light-front momentum fraction carried by the heavy quark in a hadron with momentum $P^\mu$. 
In the case of deep inelastic lepton-proton scattering, the LF momentum fraction variable $x$  in the proton structure functions can be identified with the Bjorken variable 
$x = {Q^2\over 2 p \cdot q}.$
At small $x$,  heavy-quark pairs are dominantly produced via the standard gluon-splitting subprocess $g \to Q \bar Q$. 

A recent measurement of the $c{\bar c}$ asymmetry 
has been reported by the NNPDF collaboration~\cite{NNPDF_Nov.2023}. The nonzero asymmetry
between the $D$ and ${\bar D}$ mesons extracted from  $Z+c$ production observed by the  LHCb
experiment in $pp$ collision~\cite{LHCb:2022} and an EIC experiment~\cite{EIC:2021} in $eA$ collisions can be attributed to the {\it IC} contribution in the nucleon PDF.   

In our recent article \cite{Lykasov:2025uwm}, we have confirmed the important role of the $c{\bar c}$ asymmetry
for the {\it IC} content in the proton as 
obtained from lattice gauge theory in ref.~\cite{Sufian:2020coz} and
observations~\cite{NNPDF_Nov.2023} of the $c{\bar c}$ asymmetry from
$Z+c$ production in $pp$ collisions at the LHC ~\cite{LHCb:2022}.  We show that the interference of the intrinsic $|uud c{\bar c}>$ Fock state
with the standard
contribution from the PQCD evolution leads to a large $D^+D^-$ asymmetry at
large Feynman $x$.

\section*{Acknowledgements}

Contribution to the Proceedings of  the 27th Workshop, ``What Comes Beyond
the Standard Models", Bled, Slovenia, July 8 -- 17, 2024.
I am grateful to my collaborators, including
Leonardo Di Giustino, Matin Mojaza, Xing-Gang Wu, Hung-Jung Lu, Jian-Ming Shen, Bo-Lun Du, Xu-Dong Huang, and Sheng-Quan Wang 
for their collaboration on the development and applications of the PMC, 
and
Guy  de T\'eramond, Hans Guenter Dosch, Cedric Lorc\'e, Maria Nielsen,  Tianbo Liu, Craig Roberts, Sabbir Sufian,  Philip Ratcliffe, Xing-Gang Wu, Shen Quan Wang, and Alexandre Deur, for their collaboration on light-front holography and its implications for hadron physics.
This work is supported by the Department of Energy, Contract DE--AC02--76SF00515.


\end{document}